\documentclass[prb, aps, showpacs, preprintnumbers, amsmath, amssymb, 
superscriptaddress, twocolumn]{revtex4-1}
\pdfoutput=1
\usepackage{graphicx}
\usepackage{color}
\usepackage{bm}
\usepackage{amssymb}
\usepackage{amsmath}

\begin{document}              

\title{
Structural disorder versus chiral magnetism in Cr$_{1/3}$NbS$_2$
}

\author{V. Dyadkin}
\email{diadkin@esrf.fr}
\affiliation{Swiss-Norwegian Beamlines at the European Synchrotron Radiation Facility, 38000 Grenoble, France}
\affiliation{Petersburg Nuclear Physics Institute, RNC ``Kurchatov institute'', Gatchina, 188300 St-Petersburg, Russia}

\author{F. Mushenok}
\affiliation{Institute of Problems of Chemical Physics RAS, 142432 Chernogolovka, Russia}
\affiliation{School of Physics, University of Exeter, Stocker Road, Exeter, EX4 4QL, United Kingdom}

\author{A. Bosak}
\affiliation{European Synchrotron Radiation Facility, 38000 Grenoble, France}

\author{D. Menzel}
\affiliation{Institut f\"{u}r Physik der Kondensierten Materie, Technische Universit\"{a}t Braunschweig, D-38106 Braunschweig, Germany} 

\author{S. Grigoriev}
\affiliation{Petersburg Nuclear Physics Institute, RNC ``Kurchatov institute'', Gatchina, 188300 St-Petersburg, Russia}
\affiliation{Saint-Petersburg State University, 198504 Saint-Petersburg, Russia}

\author{P. Pattison}
\affiliation{Swiss-Norwegian Beamlines at the European Synchrotron Radiation Facility, 38000 Grenoble, France}

\author{D. Chernyshov}
\affiliation{Swiss-Norwegian Beamlines at the European Synchrotron Radiation Facility, 38000 Grenoble, France}
\date{\today}

\begin{abstract}

The crystal structure of a disordered form of Cr$_{1/3}$NbS$_2$ has been characterized using diffraction and inelastic 
scattering of synchrotron radiation. In contrast to the previously reported  symmetry (P6$_3$22), the crystal can be described by a regular twinning of  an average P6$_3$ structure with three 
disordered positions of the Cr ions. Short-range correlations of the occupational disorder result in a quite intense and structured 
diffuse scattering; a static nature of the disorder was unambiguously attributed by the inelastic x-ray scattering. The diffuse 
scattering has been modeled using a reverse Monte-Carlo algorithm assuming a
disorder of the Cr sub-lattice only. 
The observed correlated disorder of the Cr sub-lattice reduces the temperature of the magnetic ordering from 130 K to 88
K and drastically modifies the field dependence of the magnetization as it is evidenced by the SQUID magnetometery. 
We conclude, that in contrast to the helicoidal spin structure assumed for P6$_3$22 form, the compound under study is ferromagnetically ordered with a pronounced in-plane anisotropy.

\end{abstract}


\maketitle

\section{Introduction}

A lack of inversion symmetry in magnetic materials makes possible the antisymmetric Dzyaloshinsky-Moriya interaction (DMI) and results in a variety of magnetic phenomena. A chiral helicoidal order \cite{BakJensen80,Maleyev06PRB}, skyrmion lattices \cite{Pfleiderer09Science,Yu2010}, a quantum blue fog \cite{PhysRevLett.96.047207}, a topological Hall effect \cite{PhysRevB.87.134424}, a first order phase transition driven by critical fluctuations \cite{PhysRevLett.105.236403} exemplify some of the topics attracting the attention of a wide range of physicists. An interesting and quite unusual circumstance is that, until recently, the most of the above listed phenomena have been proposed on the basis of the properties of MnSi and its substituted analogues. These compounds have a simple cubic crystal structure (P2$_1$3 space group) with a unit cell which contains four metal and four silicon atoms in the 4a Wyckoff positions. The crystal structure is chiral in a sense that it does not coincide with its mirror image; for the ground state, and a certain part of the $(H-T)$ (magnetic field -- temperature) phase diagram, the magnetic structure is also chiral and its chirality is unambiguously bound with the structural one \cite{Grigoriev09PRL,PhysRevB.84.014435}. Recently, the class of chiral magnets has been augmented with monogermanides of some 3d metals \cite{PhysRevB.86.134425,Dyadkin:hw5031} and Cu$_2$OSeO$_3$ \cite{PhysRevB.85.220406,PhysRevLett.108.237204,PhysRevB.89.140409}, both having the same P$2_1$3 space group as MnSi.

Another interesting example is the compound Cr$_{1/3}$NbS$_2$ \cite{Moriya1982209} where a chiral soliton lattice has been 
proposed for its magnetic structure \cite{PhysRevLett.108.107202}. Its hexagonal structure is built up from NbS$_2$ layers intercalated by Cr ions. 
There  are three basic magnetic interactions which constitute the spin structure: (i) the ferromagnetic within the Cr layers $J_\perp$, 
(ii) another ferromagnetic one $J_\parallel$ and (iii) the DMI between Cr ions, the latter two interactions belong to the two intercalating 
layers separated by NbS$_2$ \cite{JPSJ.52.1394}. 
The competition between the latter two interactions forms a helicoidal structure. 
Monte-Carlo simulations have recently shown \cite{Shinozaki2014} that the Cr ions are strongly ferromagnetically coupled in the $(ab)$-plane with $J_{\perp} \sim 74$ K, 
while they are weakly correlated between the layers with $J_\parallel \sim 10$ K and $D \sim 1.5$ K. 
The first constant $J_\perp$ determines solely the critical temperature of the compound, 
while the other two constants $D$ and $J_\parallel$ balance in the length of the helix pitch.

At variance with the crystallographically well studied B20 silicides \cite{Dmitriev2012}, the structural information on the other chiral magnets is rather incomplete and often limited to powder diffraction data. Thus, the absolute (chiral) sense has only been reported recently for both magnetism and crystal structure for Cu$_2$OSeO$_3$ \cite{PhysRevB.89.140409}, this information is not yet available for Cr$_{1/3}$NbS$_2$ because the chirality cannot be characterized from powder diffraction. As far as we are aware, the atomic coordinates in Cr$_{1/3}$NbS$_2$ were measured only once, the refinement was done using nine neutron powder peaks and it was reported without any standard deviations \cite{Hulliger1970117}. Other experimental structural information is limited to the unit cell dimensions \cite{JPSJ.52.1394}, and it is augmented with atomic coordinates borrowed from similar compounds \cite{Rouxel1971}. In spite of the great interest in unusual and potentially technologically important properties, the understanding of a microscopic nature of the chiral magnetism calls for more complete and precise information on the underlaying crystal structure.

The compounds M$_x$NbS$_2$ (where M is a transition metal) are particularly interesting due to their layered hexagonal 
structure intercalated with transition metal ions. Layered NbS$_2$--based intercalates have revealed many fascinating phenomena including 
superconductivity together with charge and spin density waves 
\cite{PhysRevLett.101.166407,PhysRevB.87.134502}, Fermi surface 
nesting effects \cite{Battaglia2007,1367-2630-10-12-125027}, and also a structural 
disorder due to a distribution of the intercalated ions over several positions. M$_x$NbS$_2$  
materials also show a variety of magnetic properties as a function of 
intercalated M and $x$:

\begin{itemize}

\item paramagnetic for intercalated Ti \cite{Hulliger1970117};

\item previously reported as paramagnetic for V \cite{Hulliger1970117}, but 
later  on as ferromagnetic below $T_c = 50$ K and another unidentified magnetic phase below $T = 20$ K with $x = 1/3$ \cite{Friend1};

\item antiferromagnetic for Co with $T_N = 26$ K and for Ni with $T_N = 90$ K 
with $x = 1/3$ \cite{Friend1,Friend2,0022-3719-16-14-016};

\item for Fe antiferromagnetic below $T_N = 150$ K with $x = 1/4$ and a spin-glass state below $T_g = 40$ K with $x = 1/3$ 
\cite{Tsuji2001213,Yamamura2004338};

\item ferromagnetic and possible helicoidal state for Mn with $x = 1/3$ below 
$T_c = 42$ and with $x = 1/4$ below $T_c = 80$ K 
\cite{Miwa1996178,Kousaka2009250};

\item helicoidal for Cr with $x = 1/3$ below $T_c = 130$ K 
\cite{Moriya1982209,Kousaka2009250}.

\end{itemize}


In spite of the interest in the magnetic properties in Cr$_{1/3}$NbS$_2$ \cite{PhysRevLett.108.107202}, there is an 
unexpectedly large spread in the reported temperatures of the magnetic ordering; 
the published values vary from 100 \cite{Mushenok2013} to 132 K 
\cite{PhysRevLett.111.197204}.

Here we revisit the intercalated crystal 
structure of the chiral helimagnet Cr$_{1/3}$NbS$_2$ using synchrotron 
radiation. All of the crystals we tested have a crystal structure which is indeed chiral but it is different from the one reported earlier. The crystal structure of Cr$_{1/3}$NbS$_2$ reveals a disorder in the Cr sub-lattice; the presence of disorder has also been confirmed in a separate diffuse scattering measurement. 
The nature of the observed diffuse component relates to short-range correlations and, as evidenced by the inelastic x-ray scattering experiment (IXS), shows the elastic peak to be the major contributor of the characteristic diffuse scattering. 

The magnetic 
susceptibility measured with SQUID magnetometery shows $T_c \approx 88$ K, such a low value points to a link between the structural disorder and magnetic properties. 
Magnetization as a function of both field and temperature also differs from the data presented earlier \cite{Kousaka2009250}, and expected from the theoretical consideration of helicoidal magnets \cite{Moriya1982209}.
Taken 
together, the new data call for a revision of theoretical models for chiral 
magnetic interactions to include effects of disorder and short range correlations in spatial 
distribution of magnetic centers. Moreover, the attribution of the given magnetic transition to the stoichiometry and disorder pattern might imply the revision of the data for other intercalates.

\section{Experimental}
\label{sec:exp}

\subsection{Sample preparation}

A polycrystalline Cr$_{1/3}$NbS$_2$ sample was prepared by sintering of the mixture of starting components at $800^\circ$ in vacuum. The Cr$_{1/3}$NbS$_2$ single crystals were grown using chemical transport method in iodine atmosphere at the temperature gradient $T = 950$ --- 800  $^\circ$C. The obtained crystals had a natural faceting and a metallic luster. The chemical composition has been independently checked with the Energy Dispersive X-Ray Analysis (EDX), with two randomly selected crystals and probing different spots. The results are very homogeneous with Cr/Nb ratio $\sim$ 0.3.

\subsection{X-ray Scattering Experiments}

Single crystal Bragg diffraction data were collected at the room temperature 
using the PILATUS@SNBL diffractometer at the BM01A end station of the 
Swiss-Norwegian Beamlines at the ESRF (Grenoble, France); the wavelength of the 
synchrotron radiation was set to 0.68290 \AA. The data were collected with a 
single $\phi$-scan with angular step of $0.1^\circ$ in a shutter-free mode with 
the PILATUS2M detector.

The raw data were processed with the SNBL Toolbox, the integral intensities 
were extracted from the frames with the CrysAlisPro software \cite{crys}, the 
crystal structure was solved with SHELXS and refined with SHELXL 
\cite{Sheldrick:sc5010}. 

Single crystal data on diffuse scattering have been measured at room temperature 
using the PILATUS6M detector at the ID23 (ESRF, Grenoble, France) with 0.689 \AA \ 
wavelength and $0.1^\circ$ slicing.
The maps of reciprocal space were recovered from the raw frames using in-house 
software with the 
orientation matrix as refined in CrysAlisPro. The diffuse scattering data 
were modeled 
with a reverse Monte-Carlo algorithm as implemented in the in-house software using 
structural models
found from the Bragg scattering experiment.

The IXS measurement was carried out at the beamline ID28 at the ESRF. The spectrometer was operated at 17.794 keV incident energy, providing an energy resolution of 3.0 meV full-width-half-maximum (Si$(9 9 9)$ reflection). IXS scans were performed in transmission geometry along selected directions in reciprocal space. Further details of the experimental set-up and the data treatment can be found elsewhere \cite{KrishBook}. 

\subsection{SQUID magnetometery}
\label{sec:squid}

The magnetization of a single crystal sample was measured using SQUID magnetometer (Quantum Design MPMS-5S). 
Temperature dependent measurements were performed in the range of $2 - 300$ K. The field dependent magnetization 
was obtained in a field range from $H = -5$ to $5$~T at $T = 6$ K. The magnetic field was applied in two different 
orientations relative to the crystallographic directions parallel to the $c$-axis and within the $ab$-plane perpendicular to $c$. 
The magnetic moment per Cr atom has been calculated from the mass and the stoichiometry of the crystal based on the EDX data. 

\section{Results}
\subsection{Magnetic properties}

\begin{figure}
 \includegraphics[width=8.5cm]{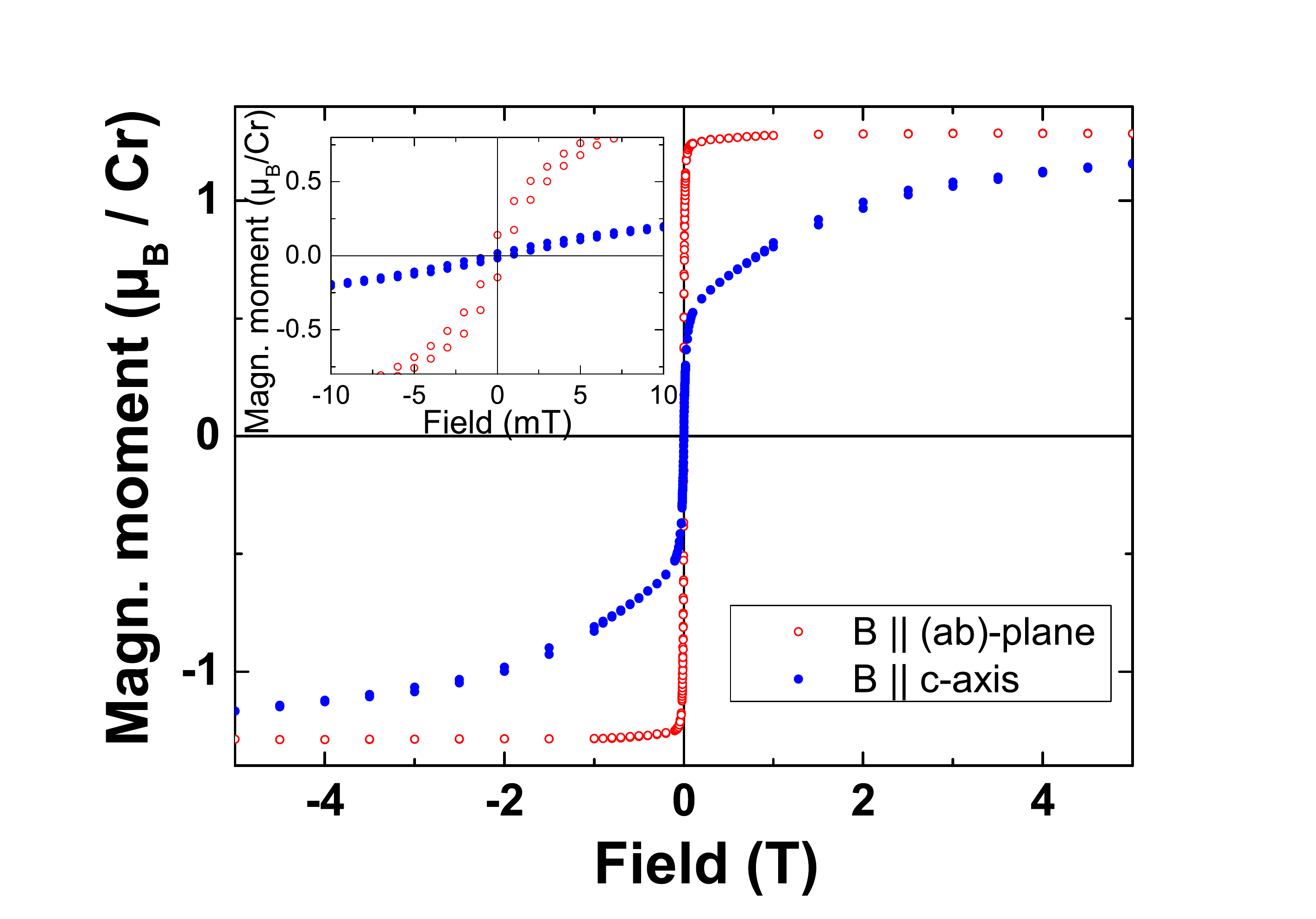}
 \caption{Magnetization curves for $H \parallel ab$ and $H \parallel c$ at $T = 6$~K.}
 \label{fig:MH}
\end{figure}

The magnetization curves at $T = 6$~K for $H \parallel c$ and $H \perp c$ show an anisotropic behavior (Fig. \ref{fig:MH}) which differs significantly from the one of the canonical CrNbS$_2$ \cite{PhysRevB.87.104403}.
For $H \parallel c$ the magnetization curve consists of two regimes: at low fields up to $H \approx 0.1$~T the magnetization increases rapidly from 0 to 0.5 $\mu_B$/Cr. At higher fields the magnetization grows smoothly up to $1.2 \mu_B$ at $5$~T. Fast saturation of magnetization for $H$ within the $ab$-plane and slow increase for $H \parallel c$ are typical features of magnetic materials with uniaxial easy-plane anisotropy. 
For $H \perp c $ the saturation field equals to 50 mT which is below the range reported previously (from 1.5 kOe \cite{Kousaka2009250} to 2 kOe \cite{PhysRevLett.108.107202}). The saturation magnetic moment $M_S = 1.3 \mu_B$ per Cr ion is considerably lower than the spin-only value of $3 \mu_B$ per Cr$^{3+}$ ions with $S = 3/2$. It is most likely that the reduction of the saturation moment is caused by delocalization of the $d$-electrons of Cr$^{3+}$.

The coercivity is smaller than 1 mT for both orientations. This value, however, is comparable with a residual magnetic field of the SQUID magnetometer and, therefore, the coercivity is below the experimental error. The negligibly small remanent magnetization and coercivity for $H \perp c$ can be explained by a continous spin rotation in the basal plane towards the external field direction. Along the $c$-axis, the field induces a canting of the spins competing against the easy-plane anisotropy.

For $H \parallel c$ a linear $M(H)$ dependence would be expected in uniaxial easy-plane ferromagnets. However, a non-linear $M(H)$ dependence is observed for the studied crystals which suggests that not only uniaxial anisotropy has to be taken into account in order to describe the magnetization behavior. Most likely, also the non-uniform distribution of the intercalated chromium ions plays an important role.

\begin{figure}
   \begin{minipage}{\linewidth}
      \includegraphics[width=8.5cm]{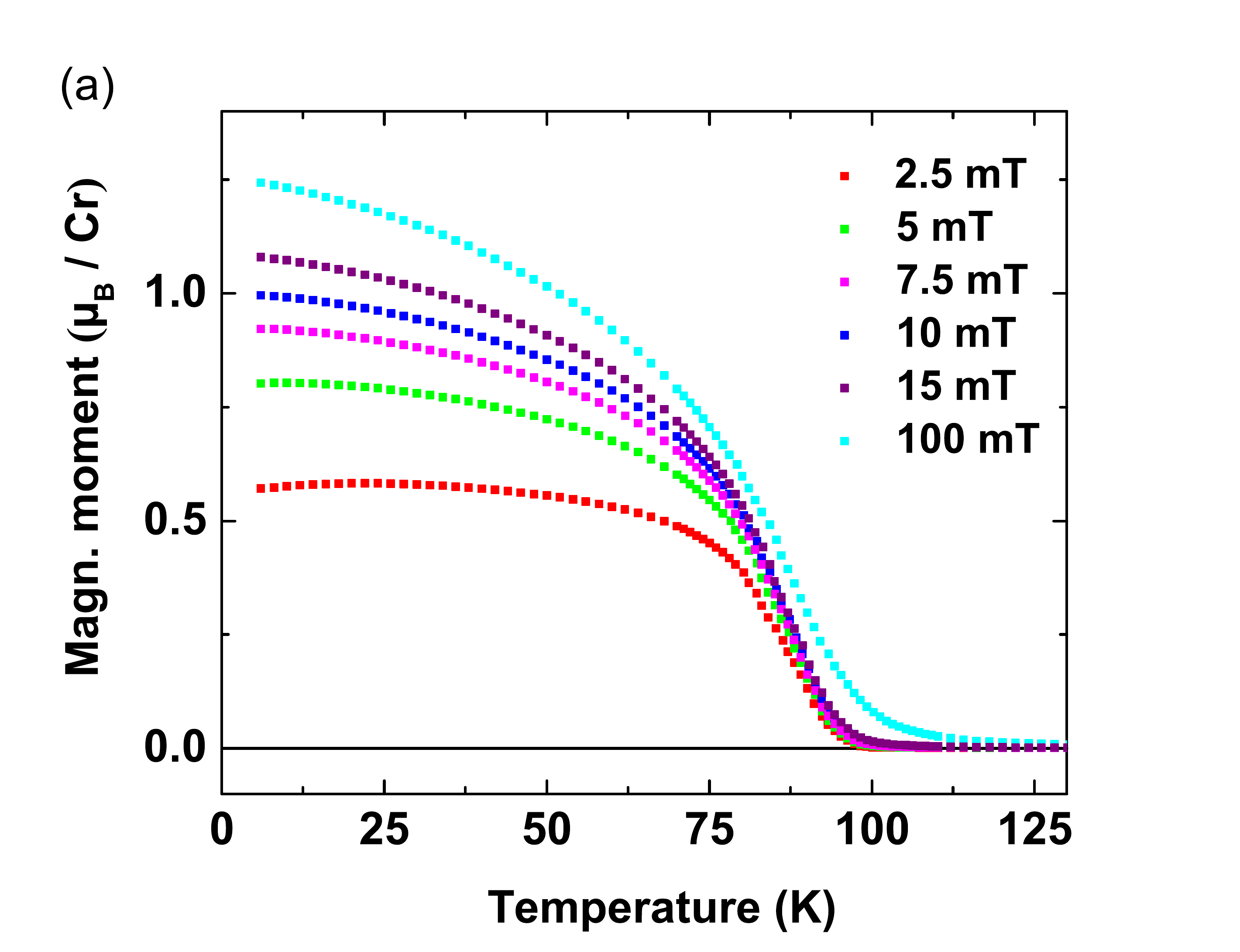}
   \end{minipage}
\\
   \begin{minipage}{\linewidth}
      \includegraphics[width=8.5cm]{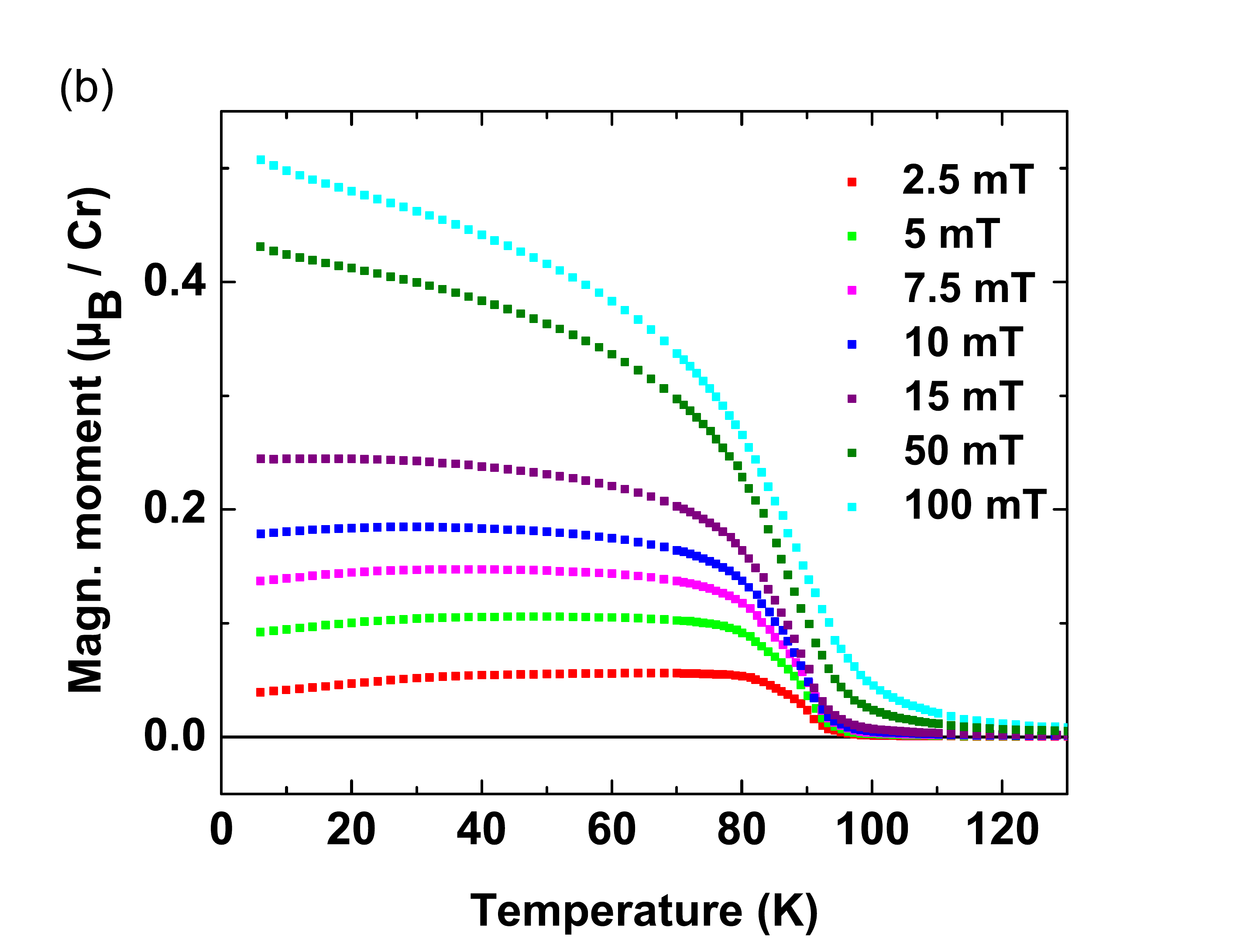}
   \end{minipage}
 \caption{Temperature dependent magnetization for (a) $H \parallel ab$ and (b) $H \parallel c$ for different magnetic fields.}
 \label{fig:MT}
\end{figure}

The temperature dependent magnetization at small magnetic fields (2.5 to 100 mT) well below saturation shows for both field orientations magnetic order below approx. $90$~K (Fig. \ref{fig:MT}). The increasing field leads to an increase of the magnetization, while the moment is smaller along the $c$-axis than within the $ab$-plane. This is also reflected by the anisotropic magnetization curves in Fig. \ref{fig:MH}.

\begin{figure}
  \begin{minipage}{\linewidth}
    \includegraphics[width=8.5cm]{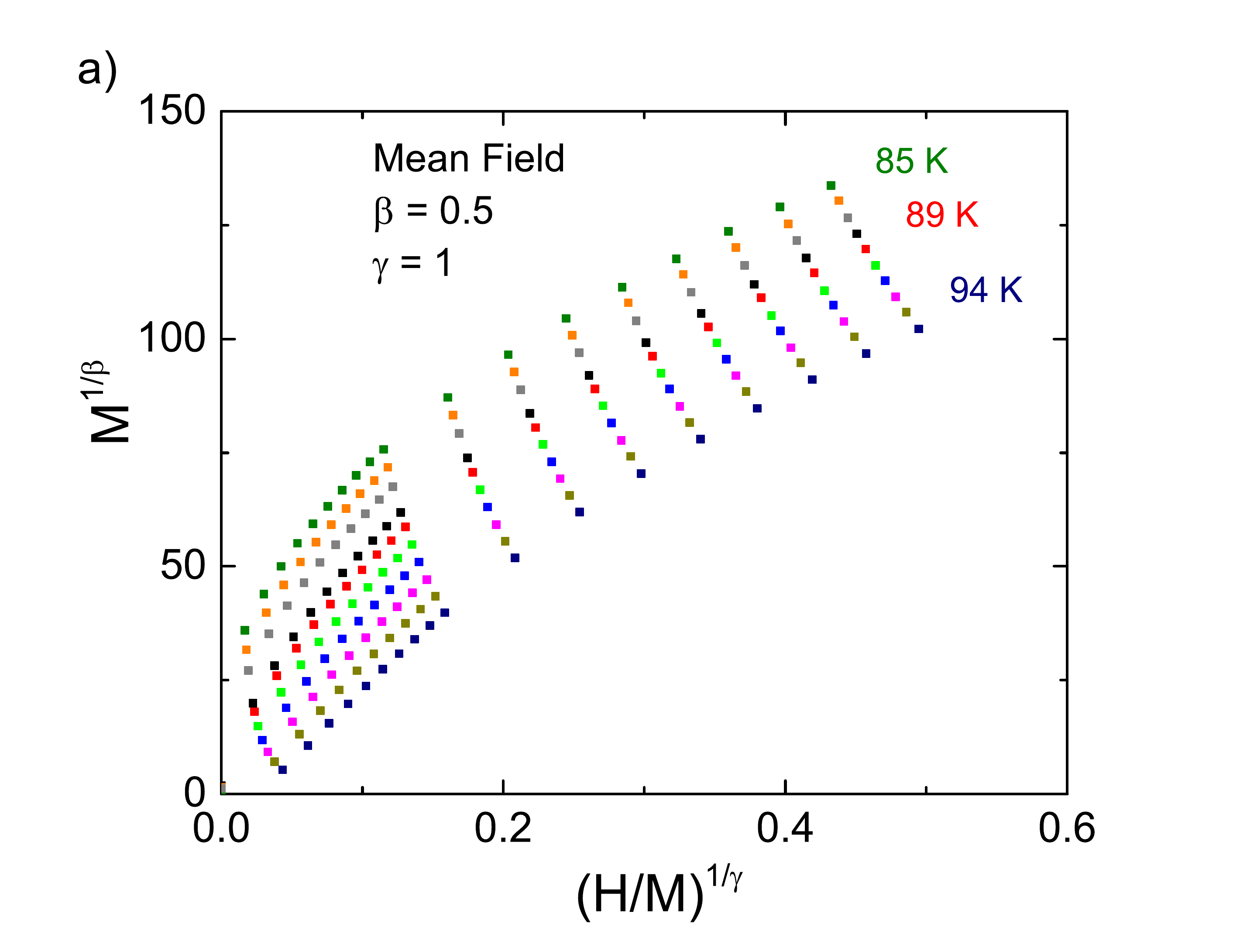} 
  \end{minipage}
  \hfill
  \begin{minipage}{\linewidth}
   \includegraphics[width=8.5cm]{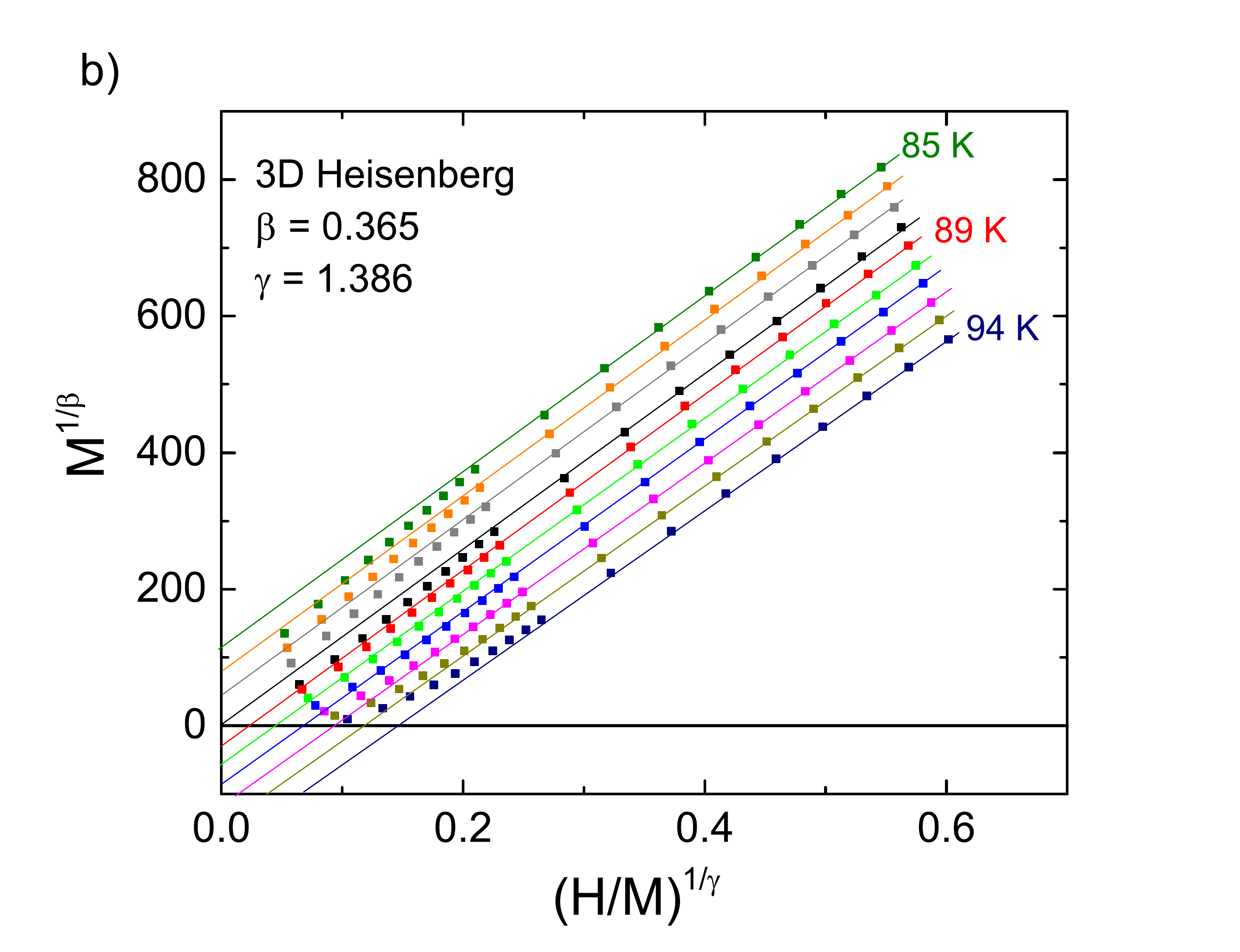} 
  \end{minipage}
\caption{Arrot plots for $H \parallel {ab}$-plane according to (a) a mean-field model ($\beta = 0.5$, $\gamma = 1$) and (b) a 3D Heisenberg model ($\beta = 0.365$, $\gamma = 1.336$).}
\label{fig:arrott}
\end{figure}

Arrott plots have been used to eliminate the influence of the external magnetic field and to accurately determine the ordering temperature. According to the mean field theory \cite{Brazovskii1975}, a plot of $M^2$ versus $H/M$ for different temperatures should show a linear dependence in the large field range. However, all curves are non-linear, but show a downward curvature (Fig. \ref{fig:arrott}a). Therefore, we modified the Arrott plot taking the critical exponents into account \cite{0034-4885-30-2-306}. A plot of $M^{1/\beta}$ versus $(H/M)^{1/\gamma}$ is shown in Fig. \ref{fig:arrott}b with the critical exponents along a 3D-Heisenberg model ($\beta = 0.365$, $\gamma = 1.336$) \cite{0295-5075-91-5-57001} representing all linear behavior at high fields. From extrapolation of the slopes we obtain an ordering temperature of $T_C = 88$~K which is significantly smaller than reported in \cite{Moriya1982209,Kousaka2009250}. The critical exponents fulfill the Widom scaling relation $\delta = 1 + \gamma / \beta$ \cite{Kadanoff1966}, which gives $\delta = 4.697$ for the experimental data when the field dependence of the magnetization is considered as $M \propto H^{1/\delta}$ at $T = T_C$. This result agrees well with $\delta = 4.797$ according to the 3D Heisenberg model in contrast to $\delta = 3$ which would hold for a mean-field model.

\subsection{Bragg diffraction and crystal structure}
\label{sec:bragg}

\begin{figure}
  \includegraphics[width=5cm,keepaspectratio]{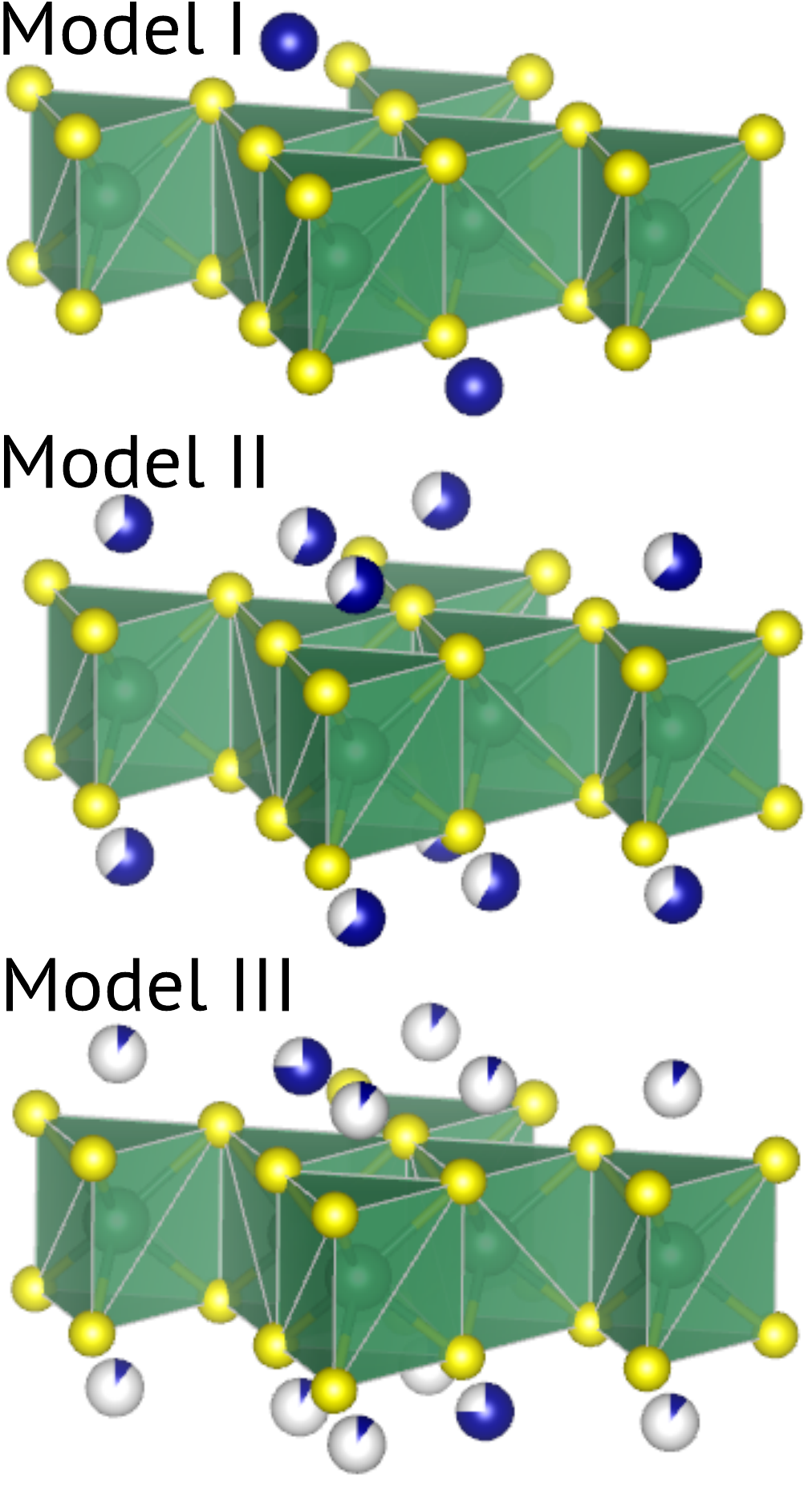}
  \caption{Fragment of crystal structures for the models I, II, III. The occupancy of disordered Cr sites is encoded as a colored sector.}
  \label{fig:models}
\end{figure}

Bragg data were indexed with a hexagonal cell ($a = 5.738(5)$ \AA, $b = 5.738(5)$ \AA, $c = 12.018(5)$ \AA) in a close agreement with the unit cell dimensions reported previously.
Three models have been tested for the crystal structure, as compared in Table 
\ref{tab:sg} and Figure \ref{fig:models}. The model I was proposed for the crystal structure of Cr$_{1/3}$NbS$_2$ with space group P6$_3$22 and  the atomic positions set close to those 
reported in Ref. \onlinecite{JPSJ.52.1394}, Cr ion occupies the $2c$ Wyckoff position.
The refinement gives rather 
high $R$-factors, and analysis of the residual electron density reveals 
additional maxima not accounted by the model. 

\begin{table}
\caption{Three possible structural models for Cr$_{1/3}$NbS$_2$. SG is the space group, WP is the Wykoff positions.}
\begin{tabular}{l|l|ll|l|l}
Model & SG & $R_1$ &$R_w$ & Cr WP & Cr/Nb Ratio \\ \hline
I & P6$_3$22 & 0.148 &0.29 & $2c$ & $0.333$, fixed by symmetry \\
II & P6$_3$22 & 0.044&0.14 & $2c$, $2b$ & $0.4$ \\
III & P6$_3$ & 0.015&0.044 & $2b$, $2a$ & $0.314$ \\
\hline
 \end{tabular}
\label{tab:sg}
\end{table}

Model II includes an additional maximum at the $2b$ Wyckoff position which indicate presence of a disordered Cr ion. Disordered M$_{1/3}$NbS$_2$ (where 
M is Mn, Fe, Co, Ni, V) has been reported, but the additional maxima were located at $2a$ and $2d$ and the
positions are found to be only slightly populated \cite{VanLaar1971154}. Refinement 
with the model II has an acceptable quality, with, however, a few warning signs:
(i) analysis of a variance of reflections included into the refinement indicates 
a possible twinning; (ii) the refined value of Cr content is  higher than the measured with EDX (the Cr/Nb ratio is expected to be $1/3$, measured is $0.3$, refined is $0.4$).

Therefore, the Model III, which has the P6$_3$ symmetry and the twinning 
law $010$ $100$ $00\bar{1}$, has been proposed. The Cr ions are found to be disordered over three independent  crystallographic positions. The refinement gives the lowest $R$-factors and the refined composition is close to the expected and measured with the Cr/Nb ratio around $0.314$. 
Corresponding crystal data and results of the refinement at the room temperature  are summarized in Tables \ref{tab:sg} and \ref{tab:cryst} and structural parameters are available as crystallographic information file (CIF) in supplementary materials.

\begin{table}
\caption{Crystal data for P6$_3$ (model III).}
\label{tab:cryst}
\begin{tabular}{l|l}
Empirical formula  & Cr$_{0.314}$NbS$_2$ \\
Temperature        & 293 K \\
Wavelength         & 0.68290 \AA \\
Crystal system, space group  &     hexagonal, P6$_3$\\ 
Unit cell dimensions      &        $a = 5.738(5)$ \AA \\
                          &        $b = 5.738(5)$ \AA  \\
                          &        $c = 12.018(5)$ \AA  \\
      Theta range for data collection &  $1.63^\circ$ to $25.25^\circ$\\
    & $-7<=h<=7$, \\
     Limiting indices     &  $-6<=k<=6$, \\
 & $-15<=l<=15$\\
   Reflections collected / unique &   2050 / 466 [R$_\mathrm{int} = 0.0198$] \\
  Completeness to theta = 25.25   &  98.8                                       
\end{tabular}
\end{table}

\subsection{Diffuse scattering and structural disorder}
\label{sec:diffuse}

\begin{figure}
  \includegraphics[width=8.5cm]{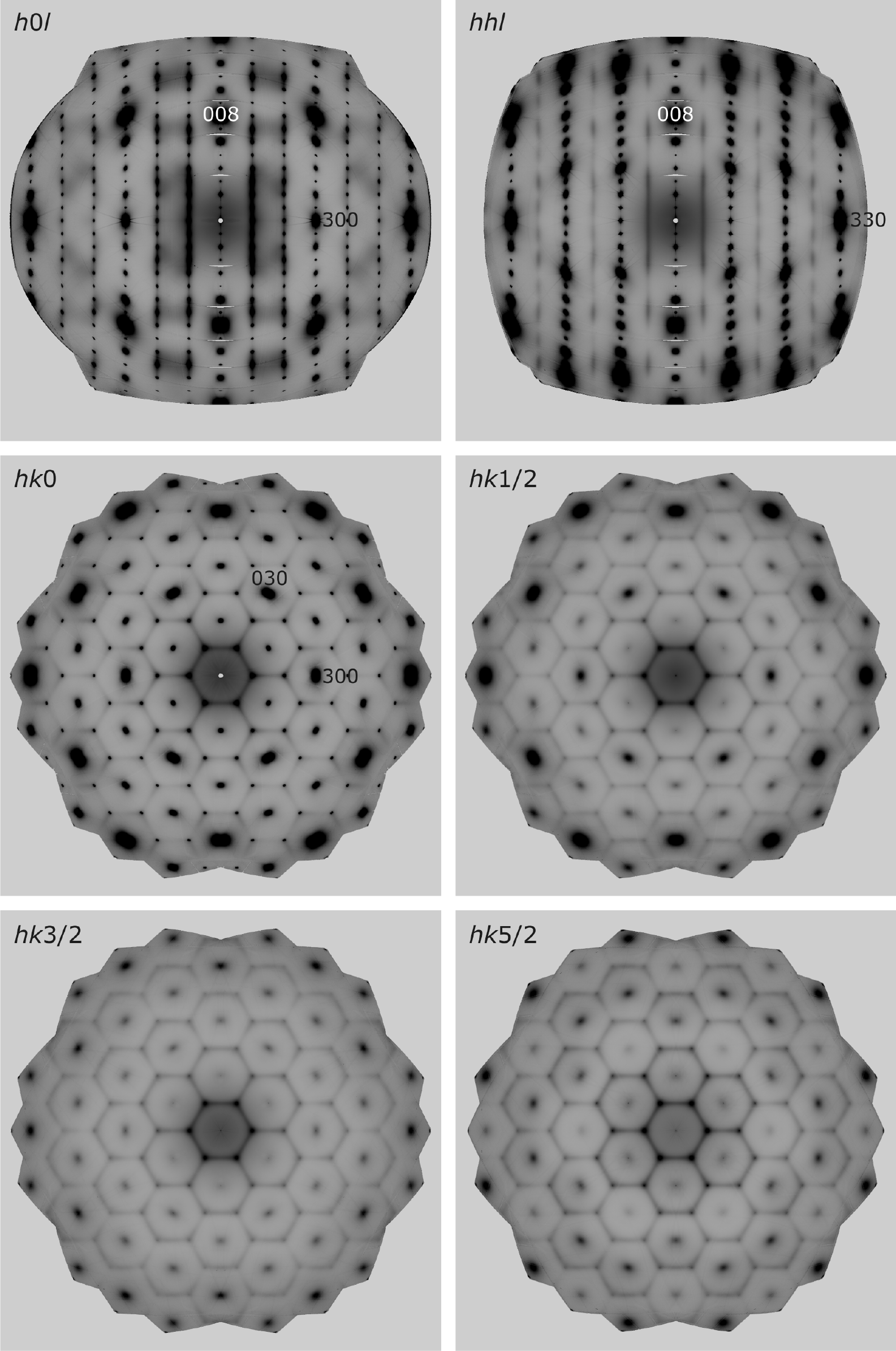}
  \caption{Maps of the diffuse scattering for different layers in reciprocal space.}
  \label{fig:diff1}
\end{figure}

Diffuse scattering has a form of a honeycomb structure extended along $c^*$, the 
walls of honeycomb cells 
are weakly modulated along this direction (Fig. \ref{fig:diff1}). 
Bragg nodes stay in the centers of honeycomb cells. Along $c^*$ the nodes are connected by the diffuse streaks which points to the presence of 2D structural defects parallel to the NbS$_2$ planes; they can be interpreted as the boundaries between the twins.

\subsection{Inelastic x-ray scattering}
\label{sec:inelastic}

\begin{figure}
  \includegraphics[width=8.5cm]{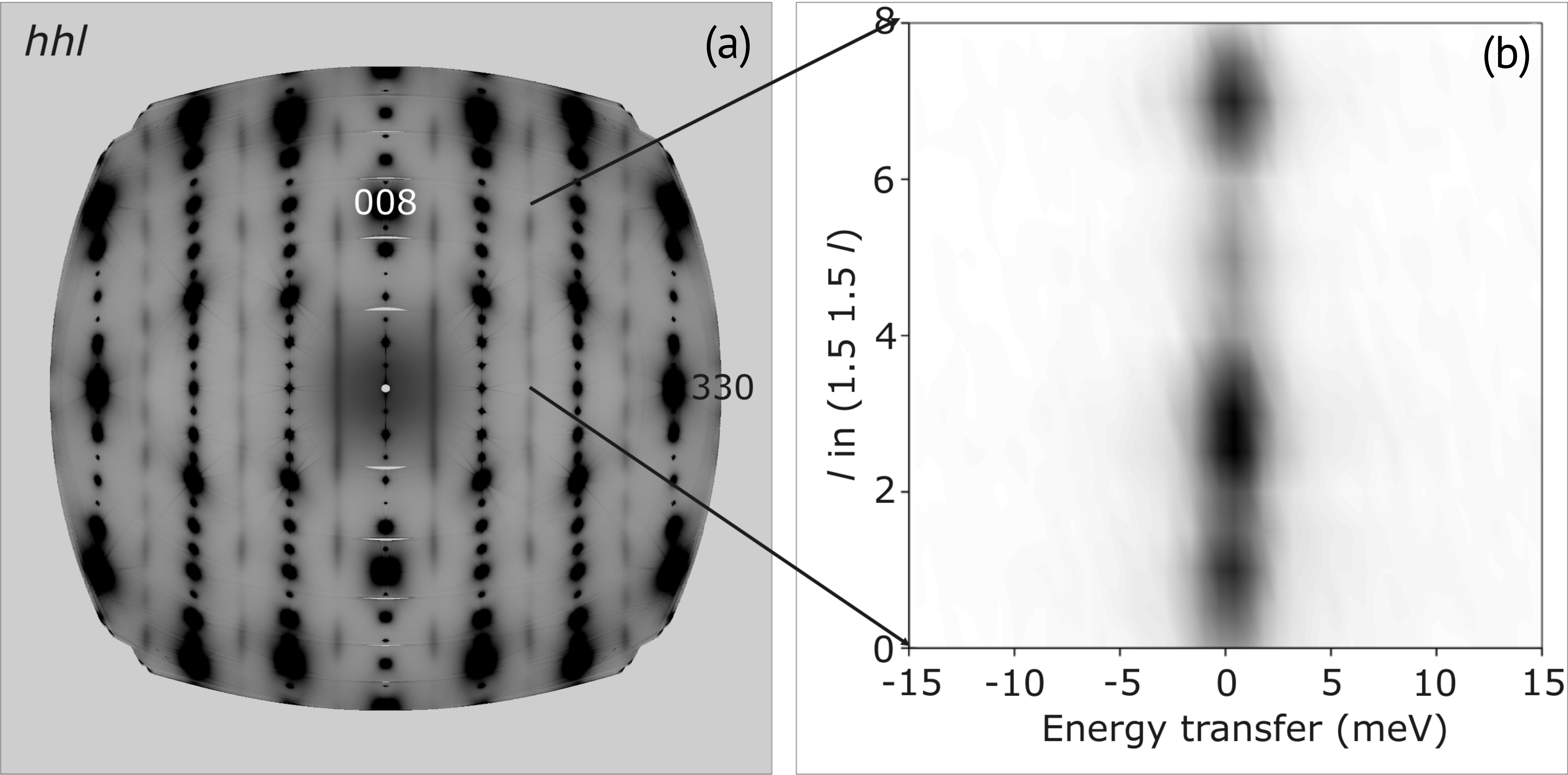}
  \caption{(a) The $hhl$-section of reciprocal space; (b) $(E,Q)$ intensity map for the region of interest indicated in (a).}
  \label{fig:diff2}
\end{figure}
Energy transfer scans along a wall of the honeycomb is shown at Fig. \ref{fig:diff2}. The intensity of the elastic peak follows the intensity of the diffuse component, while no substantial inelastic signal can be seen beyond. That 
confirms 
essentially elastic character of the observed scattering and therefore static 
nature
of the underlaying disorder.

\subsection{Monte-Carlo modeling of the disordered structure}
\label{sec:montecarlo}

Diffuse scattering has been modeled using the Reverse Monte-Carlo method (RMC) based on the Model II and using only one partially occupied Cr site (application of the twinned Model III was difficult because the number of correlating parameters was too large). 

\begin{figure}
  \includegraphics[width=8.5cm]{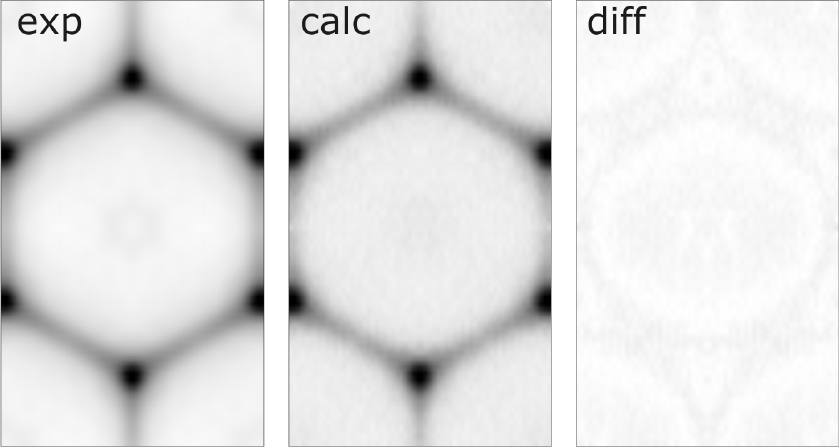}
  \caption{Diffuse scattering patterns: (exp) symmetrized experimental, (calc) resilting from the RMC, (diff) difference map. The same intensity scale is used.}
  \label{fig:diffcalc}
\end{figure}

\begin{figure}
  \includegraphics[width=8.5cm]{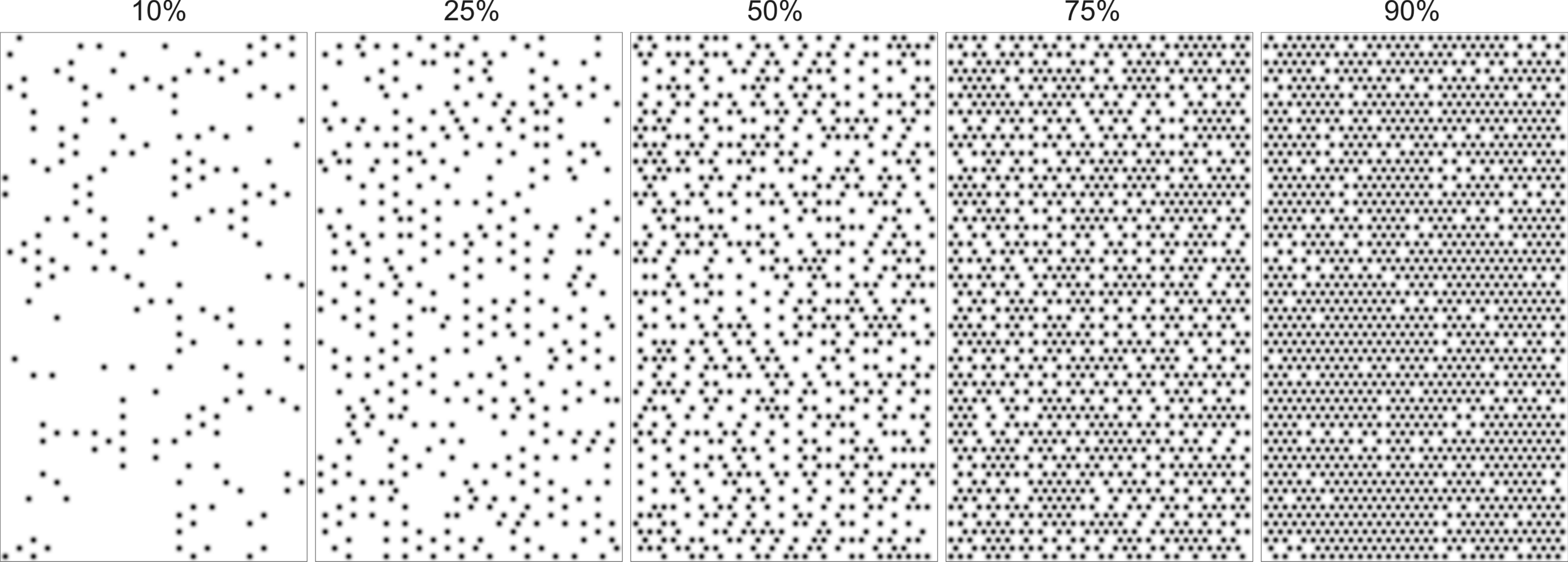}
  \caption{Randomly chosen $xy$-sections of the model clusters with the given Cr content.}
  \label{fig:montecarlo}
\end{figure}

The RMC simulation was performed on a cluster $32 \times 32 \times 32$ atoms to profit the efficiency of the fast Fourier transform. In an initial configuration the lattice sites were randomly occupied with a predefined concentration. Each simulation step included:

\begin{enumerate}
\item a swap between randomly chosen vacancy and atom;

\item calculation of the scattering intensities together with
adjustment of the background and scaling factor ($c^*$ dependence was removed by the averaging);

\item calculation of $\chi^2$ and acceptance or rejection of the swap.

\end{enumerate}

For the given cluster size, a reasonable convergence (Fig. \ref{fig:diffcalc}) is achieved in $\sim 10^5$ steps. As $c^*$ correlations were neglected, inspection of any $XY$ section of converged cluster is illustrative enough. Qualitatively, one could summarize the message as follows: vacancies/atoms are distributed in a way to minimize contacts between the neighbors of the same type, locally resulting in a loss of hexagonal packing (Fig. \ref{fig:montecarlo}).

\section{Discussion and conclusion}
\label{sec:discus}

Our results, from both Bragg and diffuse scattering unambiguously indicate that 
we are dealing with a disordered form of the title compound. Notably, the crystal 
structure of this chiral helimagnet 
 has never been accurately probed before with a single crystal diffraction 
experiment; we cannot, therefore, surely state that the disordered structure we 
observe is a generic one. However, the magnetic ordering temperature is reported inconsistently in the literature, which strongly suggests the necessity of 
a careful structural characterization in order to validate the real structure of the material under study.

 We show that the correct
symmetry is indeed chiral, but it is lower than the reported before. The P6$_3$ 
structure is twinned and disordered with respect to the Cr sub-lattice, which provides most of resonant contribution. As a result, the Flack parameter of the entire structure is not well defined. In contrast to 
 MnSi \cite{Grigoriev09PRL} and Cu$_2$OSeO$_3$ \cite{PhysRevB.89.140409} the absolute structure cannot be reliably 
determined from the crystals tested in the present study. 

The crystal structure of the new form is almost identical to the previously reported ones in terms of the geometry of the NbS$_2$ fragment. 
The main difference 
to the previously proposed structural model (P6$_3$22) is a disorder in Cr 
layers. 
In contrast to the initially expected P6$_3$22 structure that assumes only one position for the Cr ions, the experimentally observed P6$_3$ symmetry offers three Wyckoff positions with independent occupancies. Unconstrained refinement of the occupancies results in Cr$_{0.314}$NbS$_2$, that is reasonably close to the stoichiometric composition.

The disorder found in the magnetic Cr sub-lattice is not random but correlated and results in the honeycomb-like diffuse scattering shown to be essentially elastic by IXS.
This pattern suggests a tendency towards the clustering of Cr ions in hexagonal fragments within the layers. Interlayer correlations are manifested in the form of modulated diffuse rods propagating along $c^*$. Such a specific correlated disorder should strongly affect magnetic ordering, in particular, the ordering temperature; more experiments on diffuse scattering combined with magnetization measurements are needed to clarify an interplay between the correlated disorder and magnetic properties.

We conclude with a note on the necessity of further revision of crystal structure of the chiral magnets. A chiral long-range magnetic spiral (hundreds or even thousands of Angstroms) is a giant object and it should not necessarily recognize the chirality of the few Angstroms size crystal unit cell. This is the case in rare-earth magnets like Ho, where the hexagonal structure is achiral and centrosymmetric but the magnetic ordering is chiral and its chirality can be changed by an external impact \cite{PhysRevB.64.100402}. However, in the  case of chiral magnets with a chiral crystal structure,  as it was shown for B20 magnets \cite{Grigoriev10PRB1,Dyadkin20112385}, the atomic arrangement in a unit cell unambiguously dictates how the giant magnetic spiral must twist. This chiral coupling can be mapped phenomenologically by the Dzyaloshinsky-Moriya interaction, a microscopic mechanism behind it is still to be uncovered. 
Thus, the same link has been seen for both metals \cite{Grigoriev09PRL} and insulators \cite{PhysRevB.89.140409} indicating that it is not related to the conduction electrons. In the 3d-metal monogermanides the link changes its sign as a function of chemical composition, that points out the role of 3d-electrons \cite{PhysRevLett.110.207201,PhysRevB.90.174414} and, possibly, geometric factors related to the ionic radii \cite{PhysRevB.88.214402}.
The results we present here indicate that a disorder could also affect the chiral magnetism and illustrate that not only the average structure but also local deviations from it may influence the magnetic properties.

\section{Acknowledgements}
The authors are indebted to I. Snigireva (ESRF) for her help with the EDX analysis and A. Popov (ESRF) for the assistance with the diffuse scattering data collection at the ID27 beamline. We would also like to thank V. Dmitriev (SNBL) for fruitful discussions. FM was supported by the grant MK1474.2014.3 for young scientist.

\bibliography{litdb}

\begin{thebibliography}{49}%
\makeatletter
\providecommand \@ifxundefined [1]{%
 \@ifx{#1\undefined}
}%
\providecommand \@ifnum [1]{%
 \ifnum #1\expandafter \@firstoftwo
 \else \expandafter \@secondoftwo
 \fi
}%
\providecommand \@ifx [1]{%
 \ifx #1\expandafter \@firstoftwo
 \else \expandafter \@secondoftwo
 \fi
}%
\providecommand \natexlab [1]{#1}%
\providecommand \enquote  [1]{``#1''}%
\providecommand \bibnamefont  [1]{#1}%
\providecommand \bibfnamefont [1]{#1}%
\providecommand \citenamefont [1]{#1}%
\providecommand \href@noop [0]{\@secondoftwo}%
\providecommand \href [0]{\begingroup \@sanitize@url \@href}%
\providecommand \@href[1]{\@@startlink{#1}\@@href}%
\providecommand \@@href[1]{\endgroup#1\@@endlink}%
\providecommand \@sanitize@url [0]{\catcode `\\12\catcode `\$12\catcode
  `\&12\catcode `\#12\catcode `\^12\catcode `\_12\catcode `\%12\relax}%
\providecommand \@@startlink[1]{}%
\providecommand \@@endlink[0]{}%
\providecommand \url  [0]{\begingroup\@sanitize@url \@url }%
\providecommand \@url [1]{\endgroup\@href {#1}{\urlprefix }}%
\providecommand \urlprefix  [0]{URL }%
\providecommand \Eprint [0]{\href }%
\providecommand \doibase [0]{http://dx.doi.org/}%
\providecommand \selectlanguage [0]{\@gobble}%
\providecommand \bibinfo  [0]{\@secondoftwo}%
\providecommand \bibfield  [0]{\@secondoftwo}%
\providecommand \translation [1]{[#1]}%
\providecommand \BibitemOpen [0]{}%
\providecommand \bibitemStop [0]{}%
\providecommand \bibitemNoStop [0]{.\EOS\space}%
\providecommand \EOS [0]{\spacefactor3000\relax}%
\providecommand \BibitemShut  [1]{\csname bibitem#1\endcsname}%
\let\auto@bib@innerbib\@empty
\bibitem [{\citenamefont {B{\aa}k}\ and\ \citenamefont
  {Jensen}(1980)}]{BakJensen80}%
  \BibitemOpen
  \bibfield  {author} {\bibinfo {author} {\bibfnamefont {P.}~\bibnamefont
  {B{\aa}k}}\ and\ \bibinfo {author} {\bibfnamefont {M.~H.}\ \bibnamefont
  {Jensen}},\ }\href {\doibase 10.1088/0022-3719} {\bibfield  {journal}
  {\bibinfo  {journal} {J. Phys. C}\ }\textbf {\bibinfo {volume} {13}},\
  \bibinfo {pages} {L881} (\bibinfo {year} {1980})}\BibitemShut {NoStop}%
\bibitem [{\citenamefont {Maleyev}(2006)}]{Maleyev06PRB}%
  \BibitemOpen
  \bibfield  {author} {\bibinfo {author} {\bibfnamefont {S.~V.}\ \bibnamefont
  {Maleyev}},\ }\href {\doibase 10.1103/PhysRevB.73.174402} {\bibfield
  {journal} {\bibinfo  {journal} {Phys. Rev. B}\ }\textbf {\bibinfo {volume}
  {73}},\ \bibinfo {eid} {174402} (\bibinfo {year} {2006})}\BibitemShut
  {NoStop}%
\bibitem [{\citenamefont {M\"{u}hlbauer}\ \emph {et~al.}(2009)\citenamefont
  {M\"{u}hlbauer}, \citenamefont {Binz}, \citenamefont {Jonietz}, \citenamefont
  {Pfleiderer}, \citenamefont {Rosch}, \citenamefont {Neubauer}, \citenamefont
  {Georgii},\ and\ \citenamefont {B\"{o}ni}}]{Pfleiderer09Science}%
  \BibitemOpen
  \bibfield  {author} {\bibinfo {author} {\bibfnamefont {S.}~\bibnamefont
  {M\"{u}hlbauer}}, \bibinfo {author} {\bibfnamefont {B.}~\bibnamefont {Binz}},
  \bibinfo {author} {\bibfnamefont {F.}~\bibnamefont {Jonietz}}, \bibinfo
  {author} {\bibfnamefont {C.}~\bibnamefont {Pfleiderer}}, \bibinfo {author}
  {\bibfnamefont {A.}~\bibnamefont {Rosch}}, \bibinfo {author} {\bibfnamefont
  {A.}~\bibnamefont {Neubauer}}, \bibinfo {author} {\bibfnamefont
  {R.}~\bibnamefont {Georgii}}, \ and\ \bibinfo {author} {\bibfnamefont
  {P.}~\bibnamefont {B\"{o}ni}},\ }\href {\doibase 10.1126/science.1166767}
  {\bibfield  {journal} {\bibinfo  {journal} {Science}\ }\textbf {\bibinfo
  {volume} {323}},\ \bibinfo {pages} {915} (\bibinfo {year}
  {2009})}\BibitemShut {NoStop}%
\bibitem [{\citenamefont {Yu}\ \emph {et~al.}(2010)\citenamefont {Yu},
  \citenamefont {Onose}, \citenamefont {Kanazawa}, \citenamefont {Park},
  \citenamefont {Han}, \citenamefont {Matsui}, \citenamefont {Nagaosa},\ and\
  \citenamefont {Tokura}}]{Yu2010}%
  \BibitemOpen
  \bibfield  {author} {\bibinfo {author} {\bibfnamefont {X.~Z.}\ \bibnamefont
  {Yu}}, \bibinfo {author} {\bibfnamefont {Y.}~\bibnamefont {Onose}}, \bibinfo
  {author} {\bibfnamefont {N.}~\bibnamefont {Kanazawa}}, \bibinfo {author}
  {\bibfnamefont {J.~H.}\ \bibnamefont {Park}}, \bibinfo {author}
  {\bibfnamefont {J.~H.}\ \bibnamefont {Han}}, \bibinfo {author} {\bibfnamefont
  {Y.}~\bibnamefont {Matsui}}, \bibinfo {author} {\bibfnamefont
  {N.}~\bibnamefont {Nagaosa}}, \ and\ \bibinfo {author} {\bibfnamefont
  {Y.}~\bibnamefont {Tokura}},\ }\href {\doibase 10.1038/nature09124}
  {\bibfield  {journal} {\bibinfo  {journal} {Nature}\ }\textbf {\bibinfo
  {volume} {465}},\ \bibinfo {pages} {901} (\bibinfo {year}
  {2010})}\BibitemShut {NoStop}%
\bibitem [{\citenamefont {Tewari}\ \emph {et~al.}(2006)\citenamefont {Tewari},
  \citenamefont {Belitz},\ and\ \citenamefont
  {Kirkpatrick}}]{PhysRevLett.96.047207}%
  \BibitemOpen
  \bibfield  {author} {\bibinfo {author} {\bibfnamefont {S.}~\bibnamefont
  {Tewari}}, \bibinfo {author} {\bibfnamefont {D.}~\bibnamefont {Belitz}}, \
  and\ \bibinfo {author} {\bibfnamefont {T.~R.}\ \bibnamefont {Kirkpatrick}},\
  }\href {\doibase 10.1103/PhysRevLett.96.047207} {\bibfield  {journal}
  {\bibinfo  {journal} {Phys. Rev. Lett.}\ }\textbf {\bibinfo {volume} {96}},\
  \bibinfo {pages} {047207} (\bibinfo {year} {2006})}\BibitemShut {NoStop}%
\bibitem [{\citenamefont {Ritz}\ \emph {et~al.}(2013)\citenamefont {Ritz},
  \citenamefont {Halder}, \citenamefont {Franz}, \citenamefont {Bauer},
  \citenamefont {Wagner}, \citenamefont {Bamler}, \citenamefont {Rosch},\ and\
  \citenamefont {Pfleiderer}}]{PhysRevB.87.134424}%
  \BibitemOpen
  \bibfield  {author} {\bibinfo {author} {\bibfnamefont {R.}~\bibnamefont
  {Ritz}}, \bibinfo {author} {\bibfnamefont {M.}~\bibnamefont {Halder}},
  \bibinfo {author} {\bibfnamefont {C.}~\bibnamefont {Franz}}, \bibinfo
  {author} {\bibfnamefont {A.}~\bibnamefont {Bauer}}, \bibinfo {author}
  {\bibfnamefont {M.}~\bibnamefont {Wagner}}, \bibinfo {author} {\bibfnamefont
  {R.}~\bibnamefont {Bamler}}, \bibinfo {author} {\bibfnamefont
  {A.}~\bibnamefont {Rosch}}, \ and\ \bibinfo {author} {\bibfnamefont
  {C.}~\bibnamefont {Pfleiderer}},\ }\href {\doibase
  10.1103/PhysRevB.87.134424} {\bibfield  {journal} {\bibinfo  {journal} {Phys.
  Rev. B}\ }\textbf {\bibinfo {volume} {87}},\ \bibinfo {pages} {134424}
  (\bibinfo {year} {2013})}\BibitemShut {NoStop}%
\bibitem [{\citenamefont {Stishov}\ \emph {et~al.}(2010)\citenamefont
  {Stishov}, \citenamefont {Petrova}, \citenamefont {Shikov}, \citenamefont
  {Lograsso}, \citenamefont {Isaev}, \citenamefont {Johansson},\ and\
  \citenamefont {Daemen}}]{PhysRevLett.105.236403}%
  \BibitemOpen
  \bibfield  {author} {\bibinfo {author} {\bibfnamefont {S.~M.}\ \bibnamefont
  {Stishov}}, \bibinfo {author} {\bibfnamefont {A.~E.}\ \bibnamefont
  {Petrova}}, \bibinfo {author} {\bibfnamefont {A.~A.}\ \bibnamefont {Shikov}},
  \bibinfo {author} {\bibfnamefont {T.~A.}\ \bibnamefont {Lograsso}}, \bibinfo
  {author} {\bibfnamefont {E.~I.}\ \bibnamefont {Isaev}}, \bibinfo {author}
  {\bibfnamefont {B.}~\bibnamefont {Johansson}}, \ and\ \bibinfo {author}
  {\bibfnamefont {L.~L.}\ \bibnamefont {Daemen}},\ }\href {\doibase
  10.1103/PhysRevLett.105.236403} {\bibfield  {journal} {\bibinfo  {journal}
  {Phys. Rev. Lett.}\ }\textbf {\bibinfo {volume} {105}},\ \bibinfo {pages}
  {236403} (\bibinfo {year} {2010})}\BibitemShut {NoStop}%
\bibitem [{\citenamefont {Grigoriev}\ \emph {et~al.}(2009)\citenamefont
  {Grigoriev}, \citenamefont {Chernyshov}, \citenamefont {Dyadkin},
  \citenamefont {Dmitriev}, \citenamefont {Maleyev}, \citenamefont {Moskvin},
  \citenamefont {Menzel}, \citenamefont {Schoenes},\ and\ \citenamefont
  {Eckerlebe}}]{Grigoriev09PRL}%
  \BibitemOpen
  \bibfield  {author} {\bibinfo {author} {\bibfnamefont {S.~V.}\ \bibnamefont
  {Grigoriev}}, \bibinfo {author} {\bibfnamefont {D.}~\bibnamefont
  {Chernyshov}}, \bibinfo {author} {\bibfnamefont {V.~A.}\ \bibnamefont
  {Dyadkin}}, \bibinfo {author} {\bibfnamefont {V.}~\bibnamefont {Dmitriev}},
  \bibinfo {author} {\bibfnamefont {S.~V.}\ \bibnamefont {Maleyev}}, \bibinfo
  {author} {\bibfnamefont {E.~V.}\ \bibnamefont {Moskvin}}, \bibinfo {author}
  {\bibfnamefont {D.}~\bibnamefont {Menzel}}, \bibinfo {author} {\bibfnamefont
  {J.}~\bibnamefont {Schoenes}}, \ and\ \bibinfo {author} {\bibfnamefont
  {H.}~\bibnamefont {Eckerlebe}},\ }\href {\doibase
  10.1103/PhysRevLett.102.037204} {\bibfield  {journal} {\bibinfo  {journal}
  {Phys. Rev. Lett.}\ }\textbf {\bibinfo {volume} {102}},\ \bibinfo {eid}
  {037204} (\bibinfo {year} {2009})}\BibitemShut {NoStop}%
\bibitem [{\citenamefont {Dyadkin}\ \emph
  {et~al.}(2011{\natexlab{a}})\citenamefont {Dyadkin}, \citenamefont
  {Grigoriev}, \citenamefont {Menzel}, \citenamefont {Chernyshov},
  \citenamefont {Dmitriev}, \citenamefont {Schoenes}, \citenamefont {Maleyev},
  \citenamefont {Moskvin},\ and\ \citenamefont
  {Eckerlebe}}]{PhysRevB.84.014435}%
  \BibitemOpen
  \bibfield  {author} {\bibinfo {author} {\bibfnamefont {V.~A.}\ \bibnamefont
  {Dyadkin}}, \bibinfo {author} {\bibfnamefont {S.~V.}\ \bibnamefont
  {Grigoriev}}, \bibinfo {author} {\bibfnamefont {D.}~\bibnamefont {Menzel}},
  \bibinfo {author} {\bibfnamefont {D.}~\bibnamefont {Chernyshov}}, \bibinfo
  {author} {\bibfnamefont {V.}~\bibnamefont {Dmitriev}}, \bibinfo {author}
  {\bibfnamefont {J.}~\bibnamefont {Schoenes}}, \bibinfo {author}
  {\bibfnamefont {S.~V.}\ \bibnamefont {Maleyev}}, \bibinfo {author}
  {\bibfnamefont {E.~V.}\ \bibnamefont {Moskvin}}, \ and\ \bibinfo {author}
  {\bibfnamefont {H.}~\bibnamefont {Eckerlebe}},\ }\href {\doibase
  10.1103/PhysRevB.84.014435} {\bibfield  {journal} {\bibinfo  {journal} {Phys.
  Rev. B}\ }\textbf {\bibinfo {volume} {84}},\ \bibinfo {pages} {014435}
  (\bibinfo {year} {2011}{\natexlab{a}})}\BibitemShut {NoStop}%
\bibitem [{\citenamefont {Kanazawa}\ \emph {et~al.}(2012)\citenamefont
  {Kanazawa}, \citenamefont {Kim}, \citenamefont {Inosov}, \citenamefont
  {White}, \citenamefont {Egetenmeyer}, \citenamefont {Gavilano}, \citenamefont
  {Ishiwata}, \citenamefont {Onose}, \citenamefont {Arima}, \citenamefont
  {Keimer},\ and\ \citenamefont {Tokura}}]{PhysRevB.86.134425}%
  \BibitemOpen
  \bibfield  {author} {\bibinfo {author} {\bibfnamefont {N.}~\bibnamefont
  {Kanazawa}}, \bibinfo {author} {\bibfnamefont {J.-H.}\ \bibnamefont {Kim}},
  \bibinfo {author} {\bibfnamefont {D.~S.}\ \bibnamefont {Inosov}}, \bibinfo
  {author} {\bibfnamefont {J.~S.}\ \bibnamefont {White}}, \bibinfo {author}
  {\bibfnamefont {N.}~\bibnamefont {Egetenmeyer}}, \bibinfo {author}
  {\bibfnamefont {J.~L.}\ \bibnamefont {Gavilano}}, \bibinfo {author}
  {\bibfnamefont {S.}~\bibnamefont {Ishiwata}}, \bibinfo {author}
  {\bibfnamefont {Y.}~\bibnamefont {Onose}}, \bibinfo {author} {\bibfnamefont
  {T.}~\bibnamefont {Arima}}, \bibinfo {author} {\bibfnamefont
  {B.}~\bibnamefont {Keimer}}, \ and\ \bibinfo {author} {\bibfnamefont
  {Y.}~\bibnamefont {Tokura}},\ }\href {\doibase 10.1103/PhysRevB.86.134425}
  {\bibfield  {journal} {\bibinfo  {journal} {Phys. Rev. B}\ }\textbf {\bibinfo
  {volume} {86}},\ \bibinfo {pages} {134425} (\bibinfo {year}
  {2012})}\BibitemShut {NoStop}%
\bibitem [{\citenamefont {Dyadkin}\ \emph
  {et~al.}(2014{\natexlab{a}})\citenamefont {Dyadkin}, \citenamefont
  {Grigoriev}, \citenamefont {Ovsyannikov}, \citenamefont {Bykova},
  \citenamefont {Dubrovinsky}, \citenamefont {Tsvyashchenko}, \citenamefont
  {Fomicheva},\ and\ \citenamefont {Chernyshov}}]{Dyadkin:hw5031}%
  \BibitemOpen
  \bibfield  {author} {\bibinfo {author} {\bibfnamefont {V.}~\bibnamefont
  {Dyadkin}}, \bibinfo {author} {\bibfnamefont {S.}~\bibnamefont {Grigoriev}},
  \bibinfo {author} {\bibfnamefont {S.~V.}\ \bibnamefont {Ovsyannikov}},
  \bibinfo {author} {\bibfnamefont {E.}~\bibnamefont {Bykova}}, \bibinfo
  {author} {\bibfnamefont {L.}~\bibnamefont {Dubrovinsky}}, \bibinfo {author}
  {\bibfnamefont {A.}~\bibnamefont {Tsvyashchenko}}, \bibinfo {author}
  {\bibfnamefont {L.}~\bibnamefont {Fomicheva}}, \ and\ \bibinfo {author}
  {\bibfnamefont {D.}~\bibnamefont {Chernyshov}},\ }\href {\doibase
  10.1107/S2052520614006611} {\bibfield  {journal} {\bibinfo  {journal} {Acta
  Cryst. B}\ }\textbf {\bibinfo {volume} {70}},\ \bibinfo {pages} {676}
  (\bibinfo {year} {2014}{\natexlab{a}})}\BibitemShut {NoStop}%
\bibitem [{\citenamefont {Seki}\ \emph {et~al.}(2012)\citenamefont {Seki},
  \citenamefont {Kim}, \citenamefont {Inosov}, \citenamefont {Georgii},
  \citenamefont {Keimer}, \citenamefont {Ishiwata},\ and\ \citenamefont
  {Tokura}}]{PhysRevB.85.220406}%
  \BibitemOpen
  \bibfield  {author} {\bibinfo {author} {\bibfnamefont {S.}~\bibnamefont
  {Seki}}, \bibinfo {author} {\bibfnamefont {J.-H.}\ \bibnamefont {Kim}},
  \bibinfo {author} {\bibfnamefont {D.~S.}\ \bibnamefont {Inosov}}, \bibinfo
  {author} {\bibfnamefont {R.}~\bibnamefont {Georgii}}, \bibinfo {author}
  {\bibfnamefont {B.}~\bibnamefont {Keimer}}, \bibinfo {author} {\bibfnamefont
  {S.}~\bibnamefont {Ishiwata}}, \ and\ \bibinfo {author} {\bibfnamefont
  {Y.}~\bibnamefont {Tokura}},\ }\href {\doibase 10.1103/PhysRevB.85.220406}
  {\bibfield  {journal} {\bibinfo  {journal} {Phys. Rev. B}\ }\textbf {\bibinfo
  {volume} {85}},\ \bibinfo {pages} {220406} (\bibinfo {year}
  {2012})}\BibitemShut {NoStop}%
\bibitem [{\citenamefont {Adams}\ \emph {et~al.}(2012)\citenamefont {Adams},
  \citenamefont {Chacon}, \citenamefont {Wagner}, \citenamefont {Bauer},
  \citenamefont {Brandl}, \citenamefont {Pedersen}, \citenamefont {Berger},
  \citenamefont {Lemmens},\ and\ \citenamefont
  {Pfleiderer}}]{PhysRevLett.108.237204}%
  \BibitemOpen
  \bibfield  {author} {\bibinfo {author} {\bibfnamefont {T.}~\bibnamefont
  {Adams}}, \bibinfo {author} {\bibfnamefont {A.}~\bibnamefont {Chacon}},
  \bibinfo {author} {\bibfnamefont {M.}~\bibnamefont {Wagner}}, \bibinfo
  {author} {\bibfnamefont {A.}~\bibnamefont {Bauer}}, \bibinfo {author}
  {\bibfnamefont {G.}~\bibnamefont {Brandl}}, \bibinfo {author} {\bibfnamefont
  {B.}~\bibnamefont {Pedersen}}, \bibinfo {author} {\bibfnamefont
  {H.}~\bibnamefont {Berger}}, \bibinfo {author} {\bibfnamefont
  {P.}~\bibnamefont {Lemmens}}, \ and\ \bibinfo {author} {\bibfnamefont
  {C.}~\bibnamefont {Pfleiderer}},\ }\href {\doibase
  10.1103/PhysRevLett.108.237204} {\bibfield  {journal} {\bibinfo  {journal}
  {Phys. Rev. Lett.}\ }\textbf {\bibinfo {volume} {108}},\ \bibinfo {pages}
  {237204} (\bibinfo {year} {2012})}\BibitemShut {NoStop}%
\bibitem [{\citenamefont {Dyadkin}\ \emph
  {et~al.}(2014{\natexlab{b}})\citenamefont {Dyadkin}, \citenamefont
  {Pr\v{s}a}, \citenamefont {Grigoriev}, \citenamefont {White}, \citenamefont
  {Huang}, \citenamefont {R{\o}nnow}, \citenamefont {Magrez}, \citenamefont
  {Dewhurst},\ and\ \citenamefont {Chernyshov}}]{PhysRevB.89.140409}%
  \BibitemOpen
  \bibfield  {author} {\bibinfo {author} {\bibfnamefont {V.}~\bibnamefont
  {Dyadkin}}, \bibinfo {author} {\bibfnamefont {K.}~\bibnamefont {Pr\v{s}a}},
  \bibinfo {author} {\bibfnamefont {S.~V.}\ \bibnamefont {Grigoriev}}, \bibinfo
  {author} {\bibfnamefont {J.~S.}\ \bibnamefont {White}}, \bibinfo {author}
  {\bibfnamefont {P.}~\bibnamefont {Huang}}, \bibinfo {author} {\bibfnamefont
  {H.~M.}\ \bibnamefont {R{\o}nnow}}, \bibinfo {author} {\bibfnamefont
  {A.}~\bibnamefont {Magrez}}, \bibinfo {author} {\bibfnamefont {C.~D.}\
  \bibnamefont {Dewhurst}}, \ and\ \bibinfo {author} {\bibfnamefont
  {D.}~\bibnamefont {Chernyshov}},\ }\href {\doibase
  10.1103/PhysRevB.89.140409} {\bibfield  {journal} {\bibinfo  {journal} {Phys.
  Rev. B}\ }\textbf {\bibinfo {volume} {89}},\ \bibinfo {pages} {140409}
  (\bibinfo {year} {2014}{\natexlab{b}})}\BibitemShut {NoStop}%
\bibitem [{\citenamefont {Moriya}\ and\ \citenamefont
  {Miyadai}(1982)}]{Moriya1982209}%
  \BibitemOpen
  \bibfield  {author} {\bibinfo {author} {\bibfnamefont {T.}~\bibnamefont
  {Moriya}}\ and\ \bibinfo {author} {\bibfnamefont {T.}~\bibnamefont
  {Miyadai}},\ }\href {\doibase 10.1016/0038-1098(82)91006-7} {\bibfield
  {journal} {\bibinfo  {journal} {Solid State Commun.}\ }\textbf {\bibinfo
  {volume} {42}},\ \bibinfo {pages} {209} (\bibinfo {year} {1982})}\BibitemShut
  {NoStop}%
\bibitem [{\citenamefont {Togawa}\ \emph {et~al.}(2012)\citenamefont {Togawa},
  \citenamefont {Koyama}, \citenamefont {Takayanagi}, \citenamefont {Mori},
  \citenamefont {Kousaka}, \citenamefont {Akimitsu}, \citenamefont {Nishihara},
  \citenamefont {Inoue}, \citenamefont {Ovchinnikov},\ and\ \citenamefont
  {Kishine}}]{PhysRevLett.108.107202}%
  \BibitemOpen
  \bibfield  {author} {\bibinfo {author} {\bibfnamefont {Y.}~\bibnamefont
  {Togawa}}, \bibinfo {author} {\bibfnamefont {T.}~\bibnamefont {Koyama}},
  \bibinfo {author} {\bibfnamefont {K.}~\bibnamefont {Takayanagi}}, \bibinfo
  {author} {\bibfnamefont {S.}~\bibnamefont {Mori}}, \bibinfo {author}
  {\bibfnamefont {Y.}~\bibnamefont {Kousaka}}, \bibinfo {author} {\bibfnamefont
  {J.}~\bibnamefont {Akimitsu}}, \bibinfo {author} {\bibfnamefont
  {S.}~\bibnamefont {Nishihara}}, \bibinfo {author} {\bibfnamefont
  {K.}~\bibnamefont {Inoue}}, \bibinfo {author} {\bibfnamefont {A.~S.}\
  \bibnamefont {Ovchinnikov}}, \ and\ \bibinfo {author} {\bibfnamefont
  {J.}~\bibnamefont {Kishine}},\ }\href {\doibase
  10.1103/PhysRevLett.108.107202} {\bibfield  {journal} {\bibinfo  {journal}
  {Phys. Rev. Lett.}\ }\textbf {\bibinfo {volume} {108}},\ \bibinfo {pages}
  {107202} (\bibinfo {year} {2012})}\BibitemShut {NoStop}%
\bibitem [{\citenamefont {Miyadai}\ \emph {et~al.}(1983)\citenamefont
  {Miyadai}, \citenamefont {Kikuchi}, \citenamefont {Kondo}, \citenamefont
  {Sakka}, \citenamefont {Arai},\ and\ \citenamefont
  {Ishikawa}}]{JPSJ.52.1394}%
  \BibitemOpen
  \bibfield  {author} {\bibinfo {author} {\bibfnamefont {T.}~\bibnamefont
  {Miyadai}}, \bibinfo {author} {\bibfnamefont {K.}~\bibnamefont {Kikuchi}},
  \bibinfo {author} {\bibfnamefont {H.}~\bibnamefont {Kondo}}, \bibinfo
  {author} {\bibfnamefont {S.}~\bibnamefont {Sakka}}, \bibinfo {author}
  {\bibfnamefont {M.}~\bibnamefont {Arai}}, \ and\ \bibinfo {author}
  {\bibfnamefont {Y.}~\bibnamefont {Ishikawa}},\ }\href {\doibase
  10.1143/JPSJ.52.1394} {\bibfield  {journal} {\bibinfo  {journal} {J. Phys.
  Soc. Jpn.}\ }\textbf {\bibinfo {volume} {52}},\ \bibinfo {pages} {1394}
  (\bibinfo {year} {1983})}\BibitemShut {NoStop}%
\bibitem [{\citenamefont {Shinozaki}\ \emph {et~al.}(2014)\citenamefont
  {Shinozaki}, \citenamefont {Hoshino}, \citenamefont {Kishine}, \citenamefont
  {Hukushima},\ and\ \citenamefont {Kato}}]{Shinozaki2014}%
  \BibitemOpen
  \bibfield  {author} {\bibinfo {author} {\bibfnamefont {M.}~\bibnamefont
  {Shinozaki}}, \bibinfo {author} {\bibfnamefont {S.}~\bibnamefont {Hoshino}},
  \bibinfo {author} {\bibfnamefont {J.}~\bibnamefont {Kishine}}, \bibinfo
  {author} {\bibfnamefont {K.}~\bibnamefont {Hukushima}}, \ and\ \bibinfo
  {author} {\bibfnamefont {Y.}~\bibnamefont {Kato}},\ }in\ \href@noop {} {\emph
  {\bibinfo {booktitle} {Japan-Russia International Research Symposium on
  Chiral Magnetism, 6th-8th December 2014, Hiroshima (Japan)}}}\ (\bibinfo
  {year} {2014})\ p.\ \bibinfo {pages} {Abstract O04}\BibitemShut {NoStop}%
\bibitem [{\citenamefont {Dmitriev}\ \emph {et~al.}(2012)\citenamefont
  {Dmitriev}, \citenamefont {Chernyshov}, \citenamefont {Grigoriev},\ and\
  \citenamefont {Dyadkin}}]{Dmitriev2012}%
  \BibitemOpen
  \bibfield  {author} {\bibinfo {author} {\bibfnamefont {V.}~\bibnamefont
  {Dmitriev}}, \bibinfo {author} {\bibfnamefont {D.}~\bibnamefont
  {Chernyshov}}, \bibinfo {author} {\bibfnamefont {S.}~\bibnamefont
  {Grigoriev}}, \ and\ \bibinfo {author} {\bibfnamefont {V.}~\bibnamefont
  {Dyadkin}},\ }\href@noop {} {\bibfield  {journal} {\bibinfo  {journal} {J.
  Phys.: Condens. Matter}\ }\textbf {\bibinfo {volume} {24}},\ \bibinfo {pages}
  {366005} (\bibinfo {year} {2012})}\BibitemShut {NoStop}%
\bibitem [{\citenamefont {Hulliger}\ and\ \citenamefont
  {Pobitschka}(1970)}]{Hulliger1970117}%
  \BibitemOpen
  \bibfield  {author} {\bibinfo {author} {\bibfnamefont {F.}~\bibnamefont
  {Hulliger}}\ and\ \bibinfo {author} {\bibfnamefont {E.}~\bibnamefont
  {Pobitschka}},\ }\href {\doibase 10.1016/0022-4596(70)90001-0} {\bibfield
  {journal} {\bibinfo  {journal} {J. Solid State Chem.}\ }\textbf {\bibinfo
  {volume} {1}},\ \bibinfo {pages} {117} (\bibinfo {year} {1970})}\BibitemShut
  {NoStop}%
\bibitem [{\citenamefont {Rouxel}\ \emph {et~al.}(1971)\citenamefont {Rouxel},
  \citenamefont {Blanc},\ and\ \citenamefont {Royer}}]{Rouxel1971}%
  \BibitemOpen
  \bibfield  {author} {\bibinfo {author} {\bibfnamefont {J.}~\bibnamefont
  {Rouxel}}, \bibinfo {author} {\bibfnamefont {A.~L.}\ \bibnamefont {Blanc}}, \
  and\ \bibinfo {author} {\bibfnamefont {A.}~\bibnamefont {Royer}},\
  }\href@noop {} {\bibfield  {journal} {\bibinfo  {journal} {Bulletin de la
  Societe Chimique de France}\ }\textbf {\bibinfo {volume} {1971}},\ \bibinfo
  {pages} {2019} (\bibinfo {year} {1971})}\BibitemShut {NoStop}%
\bibitem [{\citenamefont {Guillam\'on}\ \emph {et~al.}(2008)\citenamefont
  {Guillam\'on}, \citenamefont {Suderow}, \citenamefont {Vieira}, \citenamefont
  {Cario}, \citenamefont {Diener},\ and\ \citenamefont
  {Rodi\`ere}}]{PhysRevLett.101.166407}%
  \BibitemOpen
  \bibfield  {author} {\bibinfo {author} {\bibfnamefont {I.}~\bibnamefont
  {Guillam\'on}}, \bibinfo {author} {\bibfnamefont {H.}~\bibnamefont
  {Suderow}}, \bibinfo {author} {\bibfnamefont {S.}~\bibnamefont {Vieira}},
  \bibinfo {author} {\bibfnamefont {L.}~\bibnamefont {Cario}}, \bibinfo
  {author} {\bibfnamefont {P.}~\bibnamefont {Diener}}, \ and\ \bibinfo {author}
  {\bibfnamefont {P.}~\bibnamefont {Rodi\`ere}},\ }\href {\doibase
  10.1103/PhysRevLett.101.166407} {\bibfield  {journal} {\bibinfo  {journal}
  {Phys. Rev. Lett.}\ }\textbf {\bibinfo {volume} {101}},\ \bibinfo {pages}
  {166407} (\bibinfo {year} {2008})}\BibitemShut {NoStop}%
\bibitem [{\citenamefont {Tissen}\ \emph {et~al.}(2013)\citenamefont {Tissen},
  \citenamefont {Osorio}, \citenamefont {Brison}, \citenamefont {Nemes},
  \citenamefont {Garc\'ia-Hern\'andez}, \citenamefont {Cario}, \citenamefont
  {Rodi\`ere}, \citenamefont {Vieira},\ and\ \citenamefont
  {Suderow}}]{PhysRevB.87.134502}%
  \BibitemOpen
  \bibfield  {author} {\bibinfo {author} {\bibfnamefont {V.~G.}\ \bibnamefont
  {Tissen}}, \bibinfo {author} {\bibfnamefont {M.~R.}\ \bibnamefont {Osorio}},
  \bibinfo {author} {\bibfnamefont {J.~P.}\ \bibnamefont {Brison}}, \bibinfo
  {author} {\bibfnamefont {N.~M.}\ \bibnamefont {Nemes}}, \bibinfo {author}
  {\bibfnamefont {M.}~\bibnamefont {Garc\'ia-Hern\'andez}}, \bibinfo {author}
  {\bibfnamefont {L.}~\bibnamefont {Cario}}, \bibinfo {author} {\bibfnamefont
  {P.}~\bibnamefont {Rodi\`ere}}, \bibinfo {author} {\bibfnamefont
  {S.}~\bibnamefont {Vieira}}, \ and\ \bibinfo {author} {\bibfnamefont
  {H.}~\bibnamefont {Suderow}},\ }\href {\doibase 10.1103/PhysRevB.87.134502}
  {\bibfield  {journal} {\bibinfo  {journal} {Phys. Rev. B}\ }\textbf {\bibinfo
  {volume} {87}},\ \bibinfo {pages} {134502} (\bibinfo {year}
  {2013})}\BibitemShut {NoStop}%
\bibitem [{\citenamefont {Battaglia}\ \emph {et~al.}(2007)\citenamefont
  {Battaglia}, \citenamefont {Cercellier}, \citenamefont {Despont},
  \citenamefont {Monney}, \citenamefont {Prester}, \citenamefont {Berger},
  \citenamefont {Forró}, \citenamefont {Garnier},\ and\ \citenamefont
  {Aebi}}]{Battaglia2007}%
  \BibitemOpen
  \bibfield  {author} {\bibinfo {author} {\bibfnamefont {C.}~\bibnamefont
  {Battaglia}}, \bibinfo {author} {\bibfnamefont {H.}~\bibnamefont
  {Cercellier}}, \bibinfo {author} {\bibfnamefont {L.}~\bibnamefont {Despont}},
  \bibinfo {author} {\bibfnamefont {C.}~\bibnamefont {Monney}}, \bibinfo
  {author} {\bibfnamefont {M.}~\bibnamefont {Prester}}, \bibinfo {author}
  {\bibfnamefont {H.}~\bibnamefont {Berger}}, \bibinfo {author} {\bibfnamefont
  {L.}~\bibnamefont {Forró}}, \bibinfo {author} {\bibfnamefont {M.~G.}\
  \bibnamefont {Garnier}}, \ and\ \bibinfo {author} {\bibfnamefont
  {P.}~\bibnamefont {Aebi}},\ }\href {\doibase 10.1140/epjb/e2007-00188-1}
  {\bibfield  {journal} {\bibinfo  {journal} {EPJ B}\ }\textbf {\bibinfo
  {volume} {57}},\ \bibinfo {pages} {385} (\bibinfo {year} {2007})}\BibitemShut
  {NoStop}%
\bibitem [{\citenamefont {Inosov}\ \emph {et~al.}(2008)\citenamefont {Inosov},
  \citenamefont {Zabolotnyy}, \citenamefont {Evtushinsky}, \citenamefont
  {Kordyuk}, \citenamefont {Büchner}, \citenamefont {Follath}, \citenamefont
  {Berger},\ and\ \citenamefont {Borisenko}}]{1367-2630-10-12-125027}%
  \BibitemOpen
  \bibfield  {author} {\bibinfo {author} {\bibfnamefont {D.~S.}\ \bibnamefont
  {Inosov}}, \bibinfo {author} {\bibfnamefont {V.~B.}\ \bibnamefont
  {Zabolotnyy}}, \bibinfo {author} {\bibfnamefont {D.~V.}\ \bibnamefont
  {Evtushinsky}}, \bibinfo {author} {\bibfnamefont {A.~A.}\ \bibnamefont
  {Kordyuk}}, \bibinfo {author} {\bibfnamefont {B.}~\bibnamefont {Büchner}},
  \bibinfo {author} {\bibfnamefont {R.}~\bibnamefont {Follath}}, \bibinfo
  {author} {\bibfnamefont {H.}~\bibnamefont {Berger}}, \ and\ \bibinfo {author}
  {\bibfnamefont {S.~V.}\ \bibnamefont {Borisenko}},\ }\href {\doibase
  10.1088/1367-2630/10/12/125027} {\bibfield  {journal} {\bibinfo  {journal}
  {New J. Phys.}\ }\textbf {\bibinfo {volume} {10}},\ \bibinfo {pages} {125027}
  (\bibinfo {year} {2008})}\BibitemShut {NoStop}%
\bibitem [{\citenamefont {Friend}\ \emph {et~al.}(1977)\citenamefont {Friend},
  \citenamefont {Beal},\ and\ \citenamefont {Yoffe}}]{Friend1}%
  \BibitemOpen
  \bibfield  {author} {\bibinfo {author} {\bibfnamefont {R.~H.}\ \bibnamefont
  {Friend}}, \bibinfo {author} {\bibfnamefont {A.~R.}\ \bibnamefont {Beal}}, \
  and\ \bibinfo {author} {\bibfnamefont {A.~D.}\ \bibnamefont {Yoffe}},\ }\href
  {\doibase 10.1080/14786437708232952} {\bibfield  {journal} {\bibinfo
  {journal} {Philos. Mag.}\ }\textbf {\bibinfo {volume} {35}},\ \bibinfo
  {pages} {1269} (\bibinfo {year} {1977})}\BibitemShut {NoStop}%
\bibitem [{\citenamefont {Parkin}\ and\ \citenamefont
  {Friend}(1980)}]{Friend2}%
  \BibitemOpen
  \bibfield  {author} {\bibinfo {author} {\bibfnamefont {S.~S.~P.}\
  \bibnamefont {Parkin}}\ and\ \bibinfo {author} {\bibfnamefont {R.~H.}\
  \bibnamefont {Friend}},\ }\href {\doibase 10.1080/13642818008245370}
  {\bibfield  {journal} {\bibinfo  {journal} {Philos. Mag. B}\ }\textbf
  {\bibinfo {volume} {41}},\ \bibinfo {pages} {65} (\bibinfo {year}
  {1980})}\BibitemShut {NoStop}%
\bibitem [{\citenamefont {Parkin}\ \emph {et~al.}(1983)\citenamefont {Parkin},
  \citenamefont {Marseglia},\ and\ \citenamefont
  {Brown}}]{0022-3719-16-14-016}%
  \BibitemOpen
  \bibfield  {author} {\bibinfo {author} {\bibfnamefont {S.~S.~P.}\
  \bibnamefont {Parkin}}, \bibinfo {author} {\bibfnamefont {E.~A.}\
  \bibnamefont {Marseglia}}, \ and\ \bibinfo {author} {\bibfnamefont {P.~J.}\
  \bibnamefont {Brown}},\ }\href@noop {} {\bibfield  {journal} {\bibinfo
  {journal} {Journal of Physics C: Solid State Physics}\ }\textbf {\bibinfo
  {volume} {16}},\ \bibinfo {pages} {2765} (\bibinfo {year}
  {1983})}\BibitemShut {NoStop}%
\bibitem [{\citenamefont {Tsuji}\ \emph {et~al.}(2001)\citenamefont {Tsuji},
  \citenamefont {Yamamura}, \citenamefont {Koyano}, \citenamefont {Katayama},\
  and\ \citenamefont {Ito}}]{Tsuji2001213}%
  \BibitemOpen
  \bibfield  {author} {\bibinfo {author} {\bibfnamefont {T.}~\bibnamefont
  {Tsuji}}, \bibinfo {author} {\bibfnamefont {Y.}~\bibnamefont {Yamamura}},
  \bibinfo {author} {\bibfnamefont {M.}~\bibnamefont {Koyano}}, \bibinfo
  {author} {\bibfnamefont {S.}~\bibnamefont {Katayama}}, \ and\ \bibinfo
  {author} {\bibfnamefont {M.}~\bibnamefont {Ito}},\ }\href {\doibase
  http://dx.doi.org/10.1016/S0925-8388(00)01329-3} {\bibfield  {journal}
  {\bibinfo  {journal} {J. Alloy. Compd.}\ }\textbf {\bibinfo {volume}
  {317–318}},\ \bibinfo {pages} {213 } (\bibinfo {year} {2001})}\BibitemShut
  {NoStop}%
\bibitem [{\citenamefont {Yamamura}\ \emph {et~al.}(2004)\citenamefont
  {Yamamura}, \citenamefont {Moriyama}, \citenamefont {Tsuji}, \citenamefont
  {Iwasa}, \citenamefont {Koyano}, \citenamefont {Katayama},\ and\
  \citenamefont {Ito}}]{Yamamura2004338}%
  \BibitemOpen
  \bibfield  {author} {\bibinfo {author} {\bibfnamefont {Y.}~\bibnamefont
  {Yamamura}}, \bibinfo {author} {\bibfnamefont {S.}~\bibnamefont {Moriyama}},
  \bibinfo {author} {\bibfnamefont {T.}~\bibnamefont {Tsuji}}, \bibinfo
  {author} {\bibfnamefont {Y.}~\bibnamefont {Iwasa}}, \bibinfo {author}
  {\bibfnamefont {M.}~\bibnamefont {Koyano}}, \bibinfo {author} {\bibfnamefont
  {S.}~\bibnamefont {Katayama}}, \ and\ \bibinfo {author} {\bibfnamefont
  {M.}~\bibnamefont {Ito}},\ }\href {\doibase
  http://dx.doi.org/10.1016/j.jallcom.2004.04.045} {\bibfield  {journal}
  {\bibinfo  {journal} {J. Alloy. Compd.}\ }\textbf {\bibinfo {volume} {383}},\
  \bibinfo {pages} {338 } (\bibinfo {year} {2004})}\BibitemShut {NoStop}%
\bibitem [{\citenamefont {Miwa}\ \emph {et~al.}(1996)\citenamefont {Miwa},
  \citenamefont {Ikuta}, \citenamefont {Hinode}, \citenamefont {Uchida},\ and\
  \citenamefont {Wakihara}}]{Miwa1996178}%
  \BibitemOpen
  \bibfield  {author} {\bibinfo {author} {\bibfnamefont {K.}~\bibnamefont
  {Miwa}}, \bibinfo {author} {\bibfnamefont {H.}~\bibnamefont {Ikuta}},
  \bibinfo {author} {\bibfnamefont {H.}~\bibnamefont {Hinode}}, \bibinfo
  {author} {\bibfnamefont {T.}~\bibnamefont {Uchida}}, \ and\ \bibinfo {author}
  {\bibfnamefont {M.}~\bibnamefont {Wakihara}},\ }\href {\doibase
  http://dx.doi.org/10.1006/jssc.1996.0282} {\bibfield  {journal} {\bibinfo
  {journal} {J. Solid State Chem.}\ }\textbf {\bibinfo {volume} {125}},\
  \bibinfo {pages} {178 } (\bibinfo {year} {1996})}\BibitemShut {NoStop}%
\bibitem [{\citenamefont {Kousaka}\ \emph {et~al.}(2009)\citenamefont
  {Kousaka}, \citenamefont {Nakao}, \citenamefont {Kishine}, \citenamefont
  {Akita}, \citenamefont {Inoue},\ and\ \citenamefont
  {Akimitsu}}]{Kousaka2009250}%
  \BibitemOpen
  \bibfield  {author} {\bibinfo {author} {\bibfnamefont {Y.}~\bibnamefont
  {Kousaka}}, \bibinfo {author} {\bibfnamefont {Y.}~\bibnamefont {Nakao}},
  \bibinfo {author} {\bibfnamefont {J.}~\bibnamefont {Kishine}}, \bibinfo
  {author} {\bibfnamefont {M.}~\bibnamefont {Akita}}, \bibinfo {author}
  {\bibfnamefont {K.}~\bibnamefont {Inoue}}, \ and\ \bibinfo {author}
  {\bibfnamefont {J.}~\bibnamefont {Akimitsu}},\ }\href {\doibase
  10.1016/j.nima.2008.11.040} {\bibfield  {journal} {\bibinfo  {journal} {Nucl.
  Instrum. Meth. A}\ }\textbf {\bibinfo {volume} {600}},\ \bibinfo {pages}
  {250} (\bibinfo {year} {2009})}\BibitemShut {NoStop}%
\bibitem [{\citenamefont {Mushenok}(2013)}]{Mushenok2013}%
  \BibitemOpen
  \bibfield  {author} {\bibinfo {author} {\bibfnamefont {F.~B.}\ \bibnamefont
  {Mushenok}},\ }\href {\doibase 10.1140/epjb/e2013-40430-7} {\bibfield
  {journal} {\bibinfo  {journal} {EPJ B}\ }\textbf {\bibinfo {volume} {86}},\
  \bibinfo {pages} {1} (\bibinfo {year} {2013})}\BibitemShut {NoStop}%
\bibitem [{\citenamefont {Togawa}\ \emph {et~al.}(2013)\citenamefont {Togawa},
  \citenamefont {Kousaka}, \citenamefont {Nishihara}, \citenamefont {Inoue},
  \citenamefont {Akimitsu}, \citenamefont {Ovchinnikov},\ and\ \citenamefont
  {Kishine}}]{PhysRevLett.111.197204}%
  \BibitemOpen
  \bibfield  {author} {\bibinfo {author} {\bibfnamefont {Y.}~\bibnamefont
  {Togawa}}, \bibinfo {author} {\bibfnamefont {Y.}~\bibnamefont {Kousaka}},
  \bibinfo {author} {\bibfnamefont {S.}~\bibnamefont {Nishihara}}, \bibinfo
  {author} {\bibfnamefont {K.}~\bibnamefont {Inoue}}, \bibinfo {author}
  {\bibfnamefont {J.}~\bibnamefont {Akimitsu}}, \bibinfo {author}
  {\bibfnamefont {A.~S.}\ \bibnamefont {Ovchinnikov}}, \ and\ \bibinfo {author}
  {\bibfnamefont {J.}~\bibnamefont {Kishine}},\ }\href {\doibase
  10.1103/PhysRevLett.111.197204} {\bibfield  {journal} {\bibinfo  {journal}
  {Phys. Rev. Lett.}\ }\textbf {\bibinfo {volume} {111}},\ \bibinfo {pages}
  {197204} (\bibinfo {year} {2013})}\BibitemShut {NoStop}%
\bibitem [{\citenamefont {{Agilent Technology}}(2013)}]{crys}%
  \BibitemOpen
  \bibfield  {author} {\bibinfo {author} {\bibnamefont {{Agilent
  Technology}}},\ }\href@noop {} {\enquote {\bibinfo {title} {{Agilent
  Technologies UK Ltd., Oxford, UK, CrysAlisPro Software system, Version
  1.171.36.28}},}\ } (\bibinfo {year} {2013})\BibitemShut {NoStop}%
\bibitem [{\citenamefont {Sheldrick}(2008)}]{Sheldrick:sc5010}%
  \BibitemOpen
  \bibfield  {author} {\bibinfo {author} {\bibfnamefont {G.~M.}\ \bibnamefont
  {Sheldrick}},\ }\href {\doibase 10.1107/S0108767307043930} {\bibfield
  {journal} {\bibinfo  {journal} {Acta Crystallogr. Sect. A}\ }\textbf
  {\bibinfo {volume} {64}},\ \bibinfo {pages} {112} (\bibinfo {year}
  {2008})}\BibitemShut {NoStop}%
\bibitem [{\citenamefont {Krisch}\ and\ \citenamefont
  {Sette}(2007)}]{KrishBook}%
  \BibitemOpen
  \bibfield  {author} {\bibinfo {author} {\bibfnamefont {M.}~\bibnamefont
  {Krisch}}\ and\ \bibinfo {author} {\bibfnamefont {F.}~\bibnamefont {Sette}},\
  }in\ \href {\doibase 10.1007/978-3-540-34436-0_5} {\emph {\bibinfo
  {booktitle} {Light Scattering in Solid IX}}},\ \bibinfo {series} {Topics in
  Applied Physics}, Vol.\ \bibinfo {volume} {108},\ \bibinfo {editor} {edited
  by\ \bibinfo {editor} {\bibfnamefont {M.}~\bibnamefont {Cardona}}\ and\
  \bibinfo {editor} {\bibfnamefont {R.}~\bibnamefont {Merlin}}}\ (\bibinfo
  {publisher} {Springer Berlin Heidelberg},\ \bibinfo {year} {2007})\ pp.\
  \bibinfo {pages} {317--370}\BibitemShut {NoStop}%
\bibitem [{\citenamefont {Ghimire}\ \emph {et~al.}(2013)\citenamefont
  {Ghimire}, \citenamefont {McGuire}, \citenamefont {Parker}, \citenamefont
  {Sipos}, \citenamefont {Tang}, \citenamefont {Yan}, \citenamefont {Sales},\
  and\ \citenamefont {Mandrus}}]{PhysRevB.87.104403}%
  \BibitemOpen
  \bibfield  {author} {\bibinfo {author} {\bibfnamefont {N.~J.}\ \bibnamefont
  {Ghimire}}, \bibinfo {author} {\bibfnamefont {M.~A.}\ \bibnamefont
  {McGuire}}, \bibinfo {author} {\bibfnamefont {D.~S.}\ \bibnamefont {Parker}},
  \bibinfo {author} {\bibfnamefont {B.}~\bibnamefont {Sipos}}, \bibinfo
  {author} {\bibfnamefont {S.}~\bibnamefont {Tang}}, \bibinfo {author}
  {\bibfnamefont {J.-Q.}\ \bibnamefont {Yan}}, \bibinfo {author} {\bibfnamefont
  {B.~C.}\ \bibnamefont {Sales}}, \ and\ \bibinfo {author} {\bibfnamefont
  {D.}~\bibnamefont {Mandrus}},\ }\href {\doibase 10.1103/PhysRevB.87.104403}
  {\bibfield  {journal} {\bibinfo  {journal} {Phys. Rev. B}\ }\textbf {\bibinfo
  {volume} {87}},\ \bibinfo {pages} {104403} (\bibinfo {year}
  {2013})}\BibitemShut {NoStop}%
\bibitem [{\citenamefont {Brazovskii}(1975)}]{Brazovskii1975}%
  \BibitemOpen
  \bibfield  {author} {\bibinfo {author} {\bibfnamefont {S.~A.}\ \bibnamefont
  {Brazovskii}},\ }\href@noop {} {\bibfield  {journal} {\bibinfo  {journal}
  {Sov. Phys. JETP}\ }\textbf {\bibinfo {volume} {41}},\ \bibinfo {pages} {85}
  (\bibinfo {year} {1975})}\BibitemShut {NoStop}%
\bibitem [{\citenamefont {Fisher}(1967)}]{0034-4885-30-2-306}%
  \BibitemOpen
  \bibfield  {author} {\bibinfo {author} {\bibfnamefont {M.~E.}\ \bibnamefont
  {Fisher}},\ }\href@noop {} {\bibfield  {journal} {\bibinfo  {journal} {Rep.
  Prog. Phys.}\ }\textbf {\bibinfo {volume} {30}},\ \bibinfo {pages} {615}
  (\bibinfo {year} {1967})}\BibitemShut {NoStop}%
\bibitem [{\citenamefont {Zhang}\ \emph {et~al.}(2010)\citenamefont {Zhang},
  \citenamefont {Fan}, \citenamefont {Li}, \citenamefont {Li}, \citenamefont
  {Ling}, \citenamefont {Qu}, \citenamefont {Tong}, \citenamefont {Tan},\ and\
  \citenamefont {Zhang}}]{0295-5075-91-5-57001}%
  \BibitemOpen
  \bibfield  {author} {\bibinfo {author} {\bibfnamefont {L.}~\bibnamefont
  {Zhang}}, \bibinfo {author} {\bibfnamefont {J.}~\bibnamefont {Fan}}, \bibinfo
  {author} {\bibfnamefont {L.}~\bibnamefont {Li}}, \bibinfo {author}
  {\bibfnamefont {R.}~\bibnamefont {Li}}, \bibinfo {author} {\bibfnamefont
  {L.}~\bibnamefont {Ling}}, \bibinfo {author} {\bibfnamefont {Z.}~\bibnamefont
  {Qu}}, \bibinfo {author} {\bibfnamefont {W.}~\bibnamefont {Tong}}, \bibinfo
  {author} {\bibfnamefont {S.}~\bibnamefont {Tan}}, \ and\ \bibinfo {author}
  {\bibfnamefont {Y.}~\bibnamefont {Zhang}},\ }\href {\doibase
  http://dx.doi.org/10.1209/0295-5075/91/57001} {\bibfield  {journal} {\bibinfo
   {journal} {EPL}\ }\textbf {\bibinfo {volume} {91}},\ \bibinfo {pages}
  {57001} (\bibinfo {year} {2010})}\BibitemShut {NoStop}%
\bibitem [{\citenamefont {Kadanoff}(1966)}]{Kadanoff1966}%
  \BibitemOpen
  \bibfield  {author} {\bibinfo {author} {\bibfnamefont {L.}~\bibnamefont
  {Kadanoff}},\ }\href@noop {} {\bibfield  {journal} {\bibinfo  {journal}
  {Physics}\ }\textbf {\bibinfo {volume} {2}},\ \bibinfo {pages} {263}
  (\bibinfo {year} {1966})}\BibitemShut {NoStop}%
\bibitem [{\citenamefont {Laar}\ \emph {et~al.}(1971)\citenamefont {Laar},
  \citenamefont {Rietveld},\ and\ \citenamefont {Ijdo}}]{VanLaar1971154}%
  \BibitemOpen
  \bibfield  {author} {\bibinfo {author} {\bibfnamefont {B.~V.}\ \bibnamefont
  {Laar}}, \bibinfo {author} {\bibfnamefont {H.}~\bibnamefont {Rietveld}}, \
  and\ \bibinfo {author} {\bibfnamefont {D.}~\bibnamefont {Ijdo}},\ }\href
  {\doibase 10.1016/0022-4596(71)90019-3} {\bibfield  {journal} {\bibinfo
  {journal} {J. Solid State Chem.}\ }\textbf {\bibinfo {volume} {3}},\ \bibinfo
  {pages} {154} (\bibinfo {year} {1971})}\BibitemShut {NoStop}%
\bibitem [{\citenamefont {Plakhty}\ \emph {et~al.}(2001)\citenamefont
  {Plakhty}, \citenamefont {Schweika}, \citenamefont {Br\"uckel}, \citenamefont
  {Kulda}, \citenamefont {Gavrilov}, \citenamefont {Regnault},\ and\
  \citenamefont {Visser}}]{PhysRevB.64.100402}%
  \BibitemOpen
  \bibfield  {author} {\bibinfo {author} {\bibfnamefont {V.~P.}\ \bibnamefont
  {Plakhty}}, \bibinfo {author} {\bibfnamefont {W.}~\bibnamefont {Schweika}},
  \bibinfo {author} {\bibfnamefont {T.}~\bibnamefont {Br\"uckel}}, \bibinfo
  {author} {\bibfnamefont {J.}~\bibnamefont {Kulda}}, \bibinfo {author}
  {\bibfnamefont {S.~V.}\ \bibnamefont {Gavrilov}}, \bibinfo {author}
  {\bibfnamefont {L.-P.}\ \bibnamefont {Regnault}}, \ and\ \bibinfo {author}
  {\bibfnamefont {D.}~\bibnamefont {Visser}},\ }\href {\doibase
  10.1103/PhysRevB.64.100402} {\bibfield  {journal} {\bibinfo  {journal} {Phys.
  Rev. B}\ }\textbf {\bibinfo {volume} {64}},\ \bibinfo {pages} {100402}
  (\bibinfo {year} {2001})}\BibitemShut {NoStop}%
\bibitem [{\citenamefont {Grigoriev}\ \emph {et~al.}(2010)\citenamefont
  {Grigoriev}, \citenamefont {Chernyshov}, \citenamefont {Dyadkin},
  \citenamefont {Dmitriev}, \citenamefont {Moskvin}, \citenamefont {Lamago},
  \citenamefont {Wolf}, \citenamefont {Menzel}, \citenamefont {Schoenes},
  \citenamefont {Maleyev},\ and\ \citenamefont {Eckerlebe}}]{Grigoriev10PRB1}%
  \BibitemOpen
  \bibfield  {author} {\bibinfo {author} {\bibfnamefont {S.~V.}\ \bibnamefont
  {Grigoriev}}, \bibinfo {author} {\bibfnamefont {D.}~\bibnamefont
  {Chernyshov}}, \bibinfo {author} {\bibfnamefont {V.~A.}\ \bibnamefont
  {Dyadkin}}, \bibinfo {author} {\bibfnamefont {V.}~\bibnamefont {Dmitriev}},
  \bibinfo {author} {\bibfnamefont {E.~V.}\ \bibnamefont {Moskvin}}, \bibinfo
  {author} {\bibfnamefont {D.}~\bibnamefont {Lamago}}, \bibinfo {author}
  {\bibfnamefont {T.}~\bibnamefont {Wolf}}, \bibinfo {author} {\bibfnamefont
  {D.}~\bibnamefont {Menzel}}, \bibinfo {author} {\bibfnamefont
  {J.}~\bibnamefont {Schoenes}}, \bibinfo {author} {\bibfnamefont {S.~V.}\
  \bibnamefont {Maleyev}}, \ and\ \bibinfo {author} {\bibfnamefont
  {H.}~\bibnamefont {Eckerlebe}},\ }\href {\doibase 10.1103/PhysRevB.81.012408}
  {\bibfield  {journal} {\bibinfo  {journal} {Phys. Rev. B}\ }\textbf {\bibinfo
  {volume} {81}},\ \bibinfo {pages} {012408} (\bibinfo {year}
  {2010})}\BibitemShut {NoStop}%
\bibitem [{\citenamefont {Dyadkin}\ \emph
  {et~al.}(2011{\natexlab{b}})\citenamefont {Dyadkin}, \citenamefont
  {Grigoriev}, \citenamefont {Menzel}, \citenamefont {Moskvin}, \citenamefont
  {Maleyev},\ and\ \citenamefont {Eckerlebe}}]{Dyadkin20112385}%
  \BibitemOpen
  \bibfield  {author} {\bibinfo {author} {\bibfnamefont {V.}~\bibnamefont
  {Dyadkin}}, \bibinfo {author} {\bibfnamefont {S.}~\bibnamefont {Grigoriev}},
  \bibinfo {author} {\bibfnamefont {D.}~\bibnamefont {Menzel}}, \bibinfo
  {author} {\bibfnamefont {E.}~\bibnamefont {Moskvin}}, \bibinfo {author}
  {\bibfnamefont {S.}~\bibnamefont {Maleyev}}, \ and\ \bibinfo {author}
  {\bibfnamefont {H.}~\bibnamefont {Eckerlebe}},\ }\href {\doibase
  10.1016/j.physb.2010.11.023} {\bibfield  {journal} {\bibinfo  {journal}
  {Physica B}\ }\textbf {\bibinfo {volume} {406}},\ \bibinfo {pages} {2385}
  (\bibinfo {year} {2011}{\natexlab{b}})}\BibitemShut {NoStop}%
\bibitem [{\citenamefont {Grigoriev}\ \emph {et~al.}(2013)\citenamefont
  {Grigoriev}, \citenamefont {Potapova}, \citenamefont {Siegfried},
  \citenamefont {Dyadkin}, \citenamefont {Moskvin}, \citenamefont {Dmitriev},
  \citenamefont {Menzel}, \citenamefont {Dewhurst}, \citenamefont {Chernyshov},
  \citenamefont {Sadykov}, \citenamefont {Fomicheva},\ and\ \citenamefont
  {Tsvyashchenko}}]{PhysRevLett.110.207201}%
  \BibitemOpen
  \bibfield  {author} {\bibinfo {author} {\bibfnamefont {S.~V.}\ \bibnamefont
  {Grigoriev}}, \bibinfo {author} {\bibfnamefont {N.~M.}\ \bibnamefont
  {Potapova}}, \bibinfo {author} {\bibfnamefont {S.-A.}\ \bibnamefont
  {Siegfried}}, \bibinfo {author} {\bibfnamefont {V.~A.}\ \bibnamefont
  {Dyadkin}}, \bibinfo {author} {\bibfnamefont {E.~V.}\ \bibnamefont
  {Moskvin}}, \bibinfo {author} {\bibfnamefont {V.}~\bibnamefont {Dmitriev}},
  \bibinfo {author} {\bibfnamefont {D.}~\bibnamefont {Menzel}}, \bibinfo
  {author} {\bibfnamefont {C.~D.}\ \bibnamefont {Dewhurst}}, \bibinfo {author}
  {\bibfnamefont {D.}~\bibnamefont {Chernyshov}}, \bibinfo {author}
  {\bibfnamefont {R.~A.}\ \bibnamefont {Sadykov}}, \bibinfo {author}
  {\bibfnamefont {L.~N.}\ \bibnamefont {Fomicheva}}, \ and\ \bibinfo {author}
  {\bibfnamefont {A.~V.}\ \bibnamefont {Tsvyashchenko}},\ }\href {\doibase
  10.1103/PhysRevLett.110.207201} {\bibfield  {journal} {\bibinfo  {journal}
  {Phys. Rev. Lett.}\ }\textbf {\bibinfo {volume} {110}},\ \bibinfo {pages}
  {207201} (\bibinfo {year} {2013})}\BibitemShut {NoStop}%
\bibitem [{\citenamefont {Grigoriev}\ \emph {et~al.}(2014)\citenamefont
  {Grigoriev}, \citenamefont {Siegfried}, \citenamefont {Altynbayev},
  \citenamefont {Potapova}, \citenamefont {Dyadkin}, \citenamefont {Moskvin},
  \citenamefont {Menzel}, \citenamefont {Heinemann}, \citenamefont {Axenov},
  \citenamefont {Fomicheva},\ and\ \citenamefont
  {Tsvyashchenko}}]{PhysRevB.90.174414}%
  \BibitemOpen
  \bibfield  {author} {\bibinfo {author} {\bibfnamefont {S.~V.}\ \bibnamefont
  {Grigoriev}}, \bibinfo {author} {\bibfnamefont {S.-A.}\ \bibnamefont
  {Siegfried}}, \bibinfo {author} {\bibfnamefont {E.~V.}\ \bibnamefont
  {Altynbayev}}, \bibinfo {author} {\bibfnamefont {N.~M.}\ \bibnamefont
  {Potapova}}, \bibinfo {author} {\bibfnamefont {V.}~\bibnamefont {Dyadkin}},
  \bibinfo {author} {\bibfnamefont {E.~V.}\ \bibnamefont {Moskvin}}, \bibinfo
  {author} {\bibfnamefont {D.}~\bibnamefont {Menzel}}, \bibinfo {author}
  {\bibfnamefont {A.}~\bibnamefont {Heinemann}}, \bibinfo {author}
  {\bibfnamefont {S.~N.}\ \bibnamefont {Axenov}}, \bibinfo {author}
  {\bibfnamefont {L.~N.}\ \bibnamefont {Fomicheva}}, \ and\ \bibinfo {author}
  {\bibfnamefont {A.~V.}\ \bibnamefont {Tsvyashchenko}},\ }\href {\doibase
  10.1103/PhysRevB.90.174414} {\bibfield  {journal} {\bibinfo  {journal} {Phys.
  Rev. B}\ }\textbf {\bibinfo {volume} {90}},\ \bibinfo {pages} {174414}
  (\bibinfo {year} {2014})}\BibitemShut {NoStop}%
\bibitem [{\citenamefont {Chizhikov}\ and\ \citenamefont
  {Dmitrienko}(2013)}]{PhysRevB.88.214402}%
  \BibitemOpen
  \bibfield  {author} {\bibinfo {author} {\bibfnamefont {V.~A.}\ \bibnamefont
  {Chizhikov}}\ and\ \bibinfo {author} {\bibfnamefont {V.~E.}\ \bibnamefont
  {Dmitrienko}},\ }\href {\doibase 10.1103/PhysRevB.88.214402} {\bibfield
  {journal} {\bibinfo  {journal} {Phys. Rev. B}\ }\textbf {\bibinfo {volume}
  {88}},\ \bibinfo {pages} {214402} (\bibinfo {year} {2013})}\BibitemShut
  {NoStop}%
\end{thebibliography}%

\end{document}